\documentclass[a4paper,12pt]{article}
\usepackage[latin1]{inputenc}
\usepackage{amsfonts}
\usepackage{amsmath}

\input{tcilatex}

\begin{document}

\title{%
\flushright{\small hep-th/0412158 \\
                   Brown preprint: BROWN-HET-1432}\\
\center{{\bf{Quantum Gauge Theories}}}\thanks{%
This review is the extended version of talks delivred by the author at Brown
university, Providence, USA 2004 and at the International School and
Workshop, Dombai, Russia 2003.}}
\author{L. Mesref\thanks{%
On leave of absence from Département d'électrotechnique, Faculté de Génie
Electrique, U.S.T.O.M.B., Oran, Algeria.}}
\date{Department of Physics, Brown University, \\
Providence, Rhode Island 02912, USA.}
\maketitle

\begin{abstract}
The scope of this review is to give a pedagogical introduction to some new
calculations and methods developed by the author in the context of quantum
groups and their applications. The review is selfcontained and serves as a
"first aid kit" before one ventures into the beautiful but bewildering
landscape of Woronowicz's theory. First, we present an up-to-date account of
the methods and definitions used in quantum gauge theories. Then, we
highlight our new results. The present paper is by no means an exhaustive
overview of this swiftly developing subject.

\bigskip

\bigskip

\bigskip

\bigskip



\bigskip

\bigskip

\bigskip

\bigskip

\bigskip

\bigskip
\end{abstract}

\tableofcontents

\newpage

\section{Introduction}

In the past two decades there has been an extremely rapid growth in the
interest of quantum groups and their applications \cite{books}. One can
mention $q$-harmonic analysis and $q$-special functions \cite{lusztig},
conformal field theories \cite{alvarez, moore, mesref1}, in the vertex and
spin models \cite{vega, pasquier}, anyons \cite{lerda, roy, ubriaco}, in
quantum optics \cite{buzek}, in the loop approach of quantum gravity \cite%
{major}, in large $N$ QCD \cite{arefeva} where the authors constructed the
master field using $q=0$ deformed commutation relations, in "fuzzy physics" 
\cite{madore} and quantum gauge theories \cite{bernard, mesref2, mesref3,
mesref4}. \newline

The quantum groups, found in the investigation of integrable systems, are a
class of noncommutative noncocomutative Hopf algebras. They were studied by
Faddeev and his collaborators \cite{faddeev1}. The initial aim of these
authors was to formulate a quantum theory of solitons \cite{takhtajan1}.
Most of their definitions are inspired by the quantum inverse scattering
method \cite{faddeev2, sklyanin, takhtajan2}. The term "quantum group" was
introduced by Drinfel'd in \cite{drinfeld}. It was considered as an
invariance group by Sudbery \cite{sudbery}. A simple example of a
noncommutative space is given by the Manin's plane \cite{manin}. In another
direction Woronowicz \cite{woronowicz}, in his seminal paper, considered
what he proposed to call pseudogroups\footnote{%
A pseudogroup $G\left( \varrho _{1},\varrho _{2},..,\varrho _{k}\right) $ of
a group $G$ is a set with a binary operation which gradually acquires the
group properties of $G$ and gradually satisfies the group axioms as some of
the parameters $\varrho _{1},\varrho _{2},..,\varrho _{k}$ of the set
approach certain limiting values or tend asymptotically to infinity. In the
case of a $q$-deformation the parameter is $\varrho =q$. In the limit $%
q\rightarrow 1$ the pseudogroup acquires the properties of a group.} and
studied bicovariant bimodules as objects analogue to tensor bundles over Lie
groups. He has also introduced the theory of bicovariant differential
calculus. This theory has turned out to be the appropriate language to study
gauge theories based on noncommutative spaces. \newline

Actually, there are maps \cite{mesref5, mesref6, mesref7,mesref8} relating
the deformed gauge fields to the ordinary ones. These maps are the analogues
of the Seiberg-Witten map \cite{seiberg}. We found these maps using the
Gerstenhaber star product \cite{gerstenhaber} instead of the
Groenewold-Moyal star product \cite{groenowold}. In ref. \cite{jevicki}, it
was proposed that quantum fluctuations in the AdS$_{3}\times $S$^{3}$
background have the effect of deforming spacetime to a noncommutative
manifold. The evidence is based on the quantum group interpretation of the
cutoff on single particle chiral primaries. In Ref. \cite{lowe}, it was
shown for the case of two-dimensional de Sitter space that there is a
natural $q$-deformation of the conformal group, with $q$ a root of unity,
where the unitary principal series representations become finite-dimensional
cyclic representations. In the framework of the dS/CFT correspondence, these
representations can lead to a description with a finite dimensional Hilbert
space and unitary evolution. The computation of the $q$-deformed metrics. of
the $q$-deformed anti-de Sitter space AdS$_{5}^{q}$ was carried out in \cite%
{mesref4}. The form of the $q$-deformed conformal correlation functions in
four dimensions was explicitly found in \cite{mesref1}. Given these results,
we hope to construct the $q$-deformed AdS$_{5}$/CFT$_{4}$ completely.
Recently, Vafa and his collaborators \cite{vafa}, have counted the number of
4-dimensional BPS black holes states on local Calabi-Yau three-folds
involving an arbitrary genus $g$ Riemann surface and showed that the
topological gauge theory on the brane reduces to a $q$-deformed $2d$
Yang-Mills theory. All these results prove that quantum groups have
unexpected applications and will certainly shed light on still open
questions in quantum gravity and quantum gauge theory. \newline

This review is organized as follows. In section 2, the concepts of quantum
groups are introduced. In section 3, we present the simple example of two
dimensional $q$-deformed plane. In section 4, we recall the celebrated
Woronowicz formalism. In section 5, we recall the $SO_{q}(6)$ bicovariant
differential calculus. In section 6, we construct the quantum gauge
transformations and the quantum BRST and anti-BRST transformations. Then, we
introduce the quantum Batalin-Vilkovisky operator. In section 7, we
introduce a map between $q$-deformed gauge fields and ordinary gauge fields.
In section 8, we study the quantum anti-de Sitter space AdS$_{5}^{q}$. We
compute the quantum metrics. and the linear transformations leading to them,
both for $q$ real and $q$ a phase. In section 9, we compute the quantum
conformal correlation functions. Section 10 is devoted to the conclusion.

\section{Quantum Groups}

Let us first consider a group $G$ in the usual sense, i.e. a set satisfying
the group axioms\footnote{%
If the invertibility condition is relaxed, we have only a semigroup.}, and $%
\mathbb{C}$ be a field of complex numbers. With this group one can associate
a commutative, associative $\mathbb{C}$-algebra of functions from $G$ to $%
\mathbb{C}$ with pointwise algebra structure, i.e. for any two elements $f$
and $f^{\prime }$, for any scalar $\alpha \in \mathbb{C}$, and $g\in G$ we
have

\begin{eqnarray}
\left( f+f^{\prime }\right) \left( g\right) &:&=f\left( g\right) +f^{\prime
}\left( g\right) ,  \notag \\
\left( \alpha f\right) \left( g\right) &:&=\alpha f\left( g\right) ,  \notag
\\
\left( f\,f^{\prime }\right) \left( g\right) &:&=f\left( g\right) f^{\prime
}\left( g\right) .
\end{eqnarray}

If $G$ is a topological group, usually only continuous functions are
considered, and for an algebraic group\footnote{%
A group $G$ is called \textbf{algebraic} if it is provided with the
structure of an algebraic variety in which the multiplication and the
inversion mappings are regular mappings (morphisms) of algebraic varieties.}
the functions are normally polynomials functions. These algebras are called
\textquotedblleft \textbf{algebras of functions on }$G$\textquotedblright . 
\newline
Example: \newline
Let $G$ be an arbitrary subgroup of the group $GL\left( N,\mathbb{C}\right) $%
. Let $Fun\left( G\right) $ be the algebra of complex valued functions on $G$%
. This algebra is unital with unit $1:G\rightarrow \mathbb{C}$, $%
g\rightarrow 1$ and is a $\ast $-algebra, where for all $f\in Fun\left(
G\right) $ the function $f^{\ast }\left( g\right) =\overline{f\left(
g\right) }$ for all $g\in G$. Consider now the matrix elements $%
M_{\,\,\,\,\,b}^{a}$ of the fundamental representation of $G$. For all $a$
and $b$, the coefficient functions: $u_{\,\,\,\,\,b}^{a}:G\rightarrow 
\mathbb{C},\quad g\rightarrow u_{\,\,\,\,\,b}^{a}\left( g\right)
=M_{\,\,\,\,\,b}^{a}$ belong to $Fun\left( G\right) $. \newline
The algebras (1) inherit some extra structures. Using the group structures
of $G$, we can introduce on the set $\mathcal{A}=Fun(G)$ of complex-valued
functions on $G$ three other linear mappings, the coproduct $\Delta $, the
counit $\epsilon $, the coinverse (or antipode) $S$: 
\begin{eqnarray}
\Delta f\left( g,g^{\prime }\right) &=&f\left( gg^{\prime }\right) ,\quad
\Delta :Fun\left( G\right) \rightarrow Fun\left( G\times G\right)  \notag \\
\epsilon \left( f\right) &=&f\left( e\right) ,\quad \quad \epsilon
:Fun\left( G\right) \rightarrow \mathbb{C},  \notag \\
S\left( f\right) \left( g\right) &=&g\left( g^{-1}\right) ,\quad S:Fun\left(
G\right) \rightarrow Fun\left( G\right)
\end{eqnarray}

where $e$ is the unit of $G$. \newline
In order to work with functions on $G$ alone, we consider the tensor product 
$Fun\left( G\right) \otimes Fun\left( G\right) $ as a linear subspace of $%
Fun\left( G\times G\right) $ by identifying \newline
$f_{1}\otimes f_{2}\in Fun\left( G\right) \otimes Fun\left( G\right) $ with
the functions $\left( f_{1}\otimes f_{2}\right) \left( g,h\right)
=f_{1}\left( g\right) f_{2}\left( h\right) $ on $G\times G$. \newline
The linear mappings satisfy the relations: \newline
\begin{eqnarray}
\left( id\otimes \Delta \right) \circ \Delta &=& \left( \Delta \otimes
id\right) \circ \Delta  \notag \\
\left( id\otimes \epsilon \right) \circ \Delta &=& \left( \epsilon \otimes
id \right) \circ \Delta =id  \notag \\
m \circ \left( S\otimes id\right) \circ \Delta &=& m\left( id\otimes
S\right) \circ \Delta = \eta \circ \epsilon
\end{eqnarray}
and 
\begin{eqnarray}
\Delta \left( ab\right) &=& \Delta \left( a\right) \Delta \left( b\right) ,
\quad \Delta \left( I \right) = I\otimes I  \notag \\
\epsilon \left( ab\right) &=& \epsilon \left( a\right) \epsilon \left( b
\right) , \quad \epsilon \left( I\right) =1  \notag \\
S\left( ab\right) &=& S\left( b\right) S\left( a\right) , \quad S\left(
I\right)=1
\end{eqnarray}
where the linear mapping (unit) $\eta : \mathbb{C}\rightarrow \mathcal{A} $
is such that $\eta \left( 1\right) $ is the unit $I$ of $\mathcal{A} $, $a,b
\in \mathcal{A} $ and $m:\mathcal{A} \rightarrow \mathcal{A} $ is the
multiplication map $m\left( a\otimes b\right) = ab$. The product in $\Delta
\left( a\right) \Delta \left( b\right) $ is the product in $\mathcal{A}
\otimes \mathcal{A} : \left( a\otimes b\right) \left( c \otimes d\right) =
ac\otimes bd$. \newline
The relations (3) and (4) define the \textbf{Hopf algebra structures} \cite%
{hopf}. \newline
For the coordinates functions $u ^{a} _{b}:G\rightarrow \mathbb{C} $ we have 
\begin{equation}
\Delta u ^{a} _{b}\left( g,h\right) = u ^{a} _{b}\left( gh\right) = \left(
gh\right) ^{a} _{b}=\sum _{c} g ^{a} _{c}h ^{c} _{b}=u ^{a} _{c}\left(
g\right) u ^{c} _{b}\left( h\right) .
\end{equation}
In general a coproduct can be expanded on $\mathcal{A} \otimes \mathcal{A} $
as: 
\begin{equation}
\Delta \left( a\right) = \sum _{i}a ^{i} _{1}\otimes a ^{i} _{2}, \quad
\quad a ^{i} _{1}, a ^{i} _{2} \in \mathcal{A} .
\end{equation}
Using Sweedler notation \cite{sweedler}, we shall suppress the index $i$ and
write this sum as: 
\begin{equation}
\Delta \left( a \right) = \sum a _{\left( 1\right) }\otimes a _{\left(
2\right) } .
\end{equation}
Here the subscripts (1), (2) refer to the corresponding tensor factors. 
\newline
Now the algebra is \textbf{deformed} or \textbf{quantized}, i.e. the algebra
structure is changed so that the algebra is not commutative anymore, but the
extra structures and axioms remain the same. This algebra is called `` 
\textbf{algebras of functions on a quantum group}'', Another definition of
quantum groups is given by:

\textit{Definition}: \newline
A quantum group is a \textbf{quasitriangular} Hopf algebra. This a pair $%
\left( \mathcal{A} , R\right) $ where $\mathcal{A} $ is a Hopf algebra and $%
R $ is an invertible element of $\mathcal{A} \otimes \mathcal{A} $ such that 
$\left( \Delta \otimes id\right) \left( R\right) = R ^{13}R ^{23}$ and $%
\Delta ^{\prime}\left( a\right) = R\Delta \left( a \right) R ^{-1}$ for $%
a\in \mathcal{A} $. \newline
Here $\Delta ^{\prime}$ is the opposite comultiplication and $R ^{12}$, $R
^{13}$, $R ^{23}$ are defined as follows: If $R=\sum _{i}x_{i}\otimes y_{i}$%
, where $x_{i},y_{i}\in \mathcal{A} $ then $R ^{12}=\sum _{i}x_{i}\otimes
y_{i}\otimes 1$, $R ^{13}=\sum _{i}x_{i}\otimes 1 \otimes y_{i}$, $R
^{23}=\sum _{i}1\otimes x_{i}\otimes y_{i}$. \newline

There are three ways of considering algebras of functions on a group and
their deformations:

(a) polynomial functions $Poly\left( G\right) $ (developed by Woronowicz and 
\newline
Drinfel'd)

(b) continuous functions $C\left( G\right) $, if $G$ is a topological group
(developed by Woronowicz)

(c) formal power series (developed by Drinfel'd). \newline

There is a similar concept of ``\textbf{quantum spaces}'': If $G$ acts on a
set $X$ (e.g. a vector space), there is a corresponding so-called \textbf{\
coaction} of the commutative algebra of functions on $G$ on the commutative
algebra of functions on $X$ satisfying certain axioms. The latter algebra
can often be deformed into a non-commutative algebra called the ``\textbf{\
algebra of functions on a quantum space}''. \newline

If we consider a compact Hausdorff space $X$, and the set $C\left( X\right) $
of continuous, complex valued functions on $X$. $C\left( X\right) $ is
naturally endowed with the structure of a commutative algebra with unit over
the complex number field, equipped moreover with anti-linear involution $%
\ast $ given by

\begin{equation}
\left( f^{\ast }\right) \left( p\right) =\overline{f\left( p\right) }
\end{equation}

and the norm

\begin{equation}
\left\| f\right\| =\sup_{p\in X}\left| f\left( p\right) \right| .
\end{equation}

This norm can be seen to obey the condition

\begin{equation}
\left\| f^{\ast }f\right\| =\left\| f\right\| ^{2}.
\end{equation}

Algebras with the above properties are known as $C^{\ast }$-algebras. We see
therefore that every compact Hausdorff space $X$ is in a natural way
associated with a commutative $C^{\ast }$-algebra with unit\footnote{%
Every commutative $C^{\star }$-algebra may be identified with an algebra of
continuous functions on a locally compact topological space. If the algebra
has a unit element, this space is moreover compact. In the contrary case, we
are dealing with the algebra of continuous functions on noncompact space,
subject to the condition of vanishing at infinity.}, namely $C\left(
X\right) $. Moreover, every continuous mapping between compact Hausdorff
spaces, $T:X\rightarrow Y$, determines a $C^{\ast }$-homomorphism: $T^{\ast
}:C\left( Y\right) \rightarrow C\left( X\right) $, given by

\begin{equation}
\left( T^{\ast }f\right) \left( p\right) =f\left( T\left( p\right) \right) .
\end{equation}

Points of $Y$ correspond to linear multiplicative functionals on $C\left(
Y\right) $. The \textbf{Gelfand-Naimark} theorem \cite{gelfand} states that
this correspondence is one to one\footnote{%
In the language of category theory: there exists a contravariant isomorphism
between the category of compact topological spaces and that of $C^{\star }$
-algebras with unit.}. Consequently, it is natural to make the following
generalization: A compact quantum space corresponds, by an extension of this
isomorphism, to a noncommutative $C^{\star }$-algebra with unit.

\section{Manin's Construction}

A simple example of a quantum space is given by the Manin's plane\footnote{%
The quantum plane approach was first suggested by Yu Kobyzev (Moscow, winter
1986 and developed by Manin at université de Montréal in June 1988 .}. The
quantum plane $R_{q}\left[ 2,0\right] $ is defined, according to Manin \cite%
{manin}, in terms of two variables $\hat{x},\,\hat{y}$, which satisfy the
commutation relations

\begin{equation}
\hat{x}\hat{y}=q\hat{y}\hat{x}
\end{equation}

where $q$ is a complex number\footnote{%
Manin uses $q^{-1}$ where we use $q$, following the usage of the Leningrad
school \cite{faddeev1}.}. The coordinates neither commute nor anticommute
unless $q=\pm 1$, respectively. Now consider a matrix

\begin{equation}
M=\left( 
\begin{array}{cc}
a & b \\ 
c & d%
\end{array}
\right) \in GL_{q}\left( 2\right)
\end{equation}

such that

\begin{eqnarray}
\hat{x}^{\prime } &=&a\hat{x}+b\hat{y}  \notag \\
\hat{y}^{\prime } &=&c\hat{x}+d\hat{y}
\end{eqnarray}

and $\left( \hat{x}^{\prime },\hat{y}^{\prime }\right) \in R_{q}\left[ 2,0%
\right] $. The elements of $M$ are supposed to commute with $x,y$. This
condition imposes restrictions upon $M$, giving the $GL_{q}\left( 2\right) $
relations

\begin{eqnarray}
ab &=&qba,\qquad cd=qdc,  \notag \\
ac &=&qca,\qquad bc=cb,  \notag \\
bd &=&qdb,\qquad ad-da=\left( q-q^{-1}\right) bc.
\end{eqnarray}

The classical case is obtained by setting $q$ equal to one.

Using these relations, it is easy to show that $D_{q}=ad-qbc$ commutes with
all the elements $a,b,c,d$ and thus may be considered as a number, the
\textquotedblleft quantum determinant\footnote{%
The quantum determinant is defined as the determinant of the middle matrix
in the Borel decomposition:
\par
$\left( 
\begin{array}{cc}
a & b \\ 
c & d%
\end{array}%
\right) =\left( 
\begin{array}{cc}
1 & bd^{-1} \\ 
0 & 1%
\end{array}%
\right) \left( 
\begin{array}{cc}
a-bd^{-1}c & 0 \\ 
0 & d%
\end{array}%
\right) \left( 
\begin{array}{cc}
1 & 0 \\ 
d^{-1}c & 1%
\end{array}%
\right) .$%
\par
{}}\textquotedblright . The choice $D_{q}=1$ restricts the quantum group to $%
SL_{q}\left( 2\right) $. Because $D_{q}$ commutes with elements of $M$ there
exists an inverse

\begin{equation}
M^{-1}=\left( D_{q}\right) ^{-1}\left( 
\begin{array}{cc}
d & -q^{-1}b \\ 
-qc & a%
\end{array}
\right) ,
\end{equation}

which is both a left and right inverse for $M$. Note that $M^{-1}$ is a
member of $GL_{q^{-1}}\left( 2\right) $ rather than $GL_{q}\left( 2\right) $
, and thus $GL_{q}\left( 2\right) $ is not, strictly speaking, a group. The
algebra (15) is associative under multiplication and the relations may be
reexpressed in a tensor form

\begin{equation}
R_{\,\,\,\,\,kl}^{ij}M_{\,\,\,m}^{k}M_{\,\,\,n}^{l}=M_{\,\,\,k}^{i}M_{\,\,
\,l}^{j}R_{\,\,\,\,\,mn}^{kl}
\end{equation}

where $R_{\,\,\,\,\,kl}^{ij}=R_{\,\,\,\,\,kl}^{ij}\,\left( q\right) $ is a
braiding matrix, whose explicit form is given by\footnote{%
Castellani \cite{books} and other authors use the following relation
\par
$\mathcal{R}_{\,\,\,\,\,\,kl}^{ij}\left( q\right)
M_{\,\,\,m}^{k}M_{\,\,\,n}^{l}=M_{\,\,\,k}^{i}M_{\,\,\,l}^{j}\mathcal{R}
_{\,\,\,\,\,mn}^{lk}\left( q\right) $ where $\mathcal{R}_{\,\,\,\,\,
\,kl}^{ij}\left( q\right) =\left( 
\begin{array}{cccc}
q & 0 & 0 & 0 \\ 
0 & 1 & 0 & 0 \\ 
0 & q-q^{-1} & 1 & 0 \\ 
0 & 0 & 0 & q%
\end{array}
\right) $.
\par
This matrix appeared for the first time in the paper \cite{faddeev3}. The
relation between the two matrices is given by $\mathcal{R}
_{\,\,\,\,\,\,kl}^{ij}\left( q\right) =qR_{\,\,\,\,\,lk}^{ij}$ $\left(
q^{-1}\right) = q \left( R ^{-1}\right) ^{ij} _{kl}\left( q\right) $.
\par
The matrix $\mathcal{R}_{\,\,\,\,\,\,kl}^{ij}$ satisfies the well known
Yang-Baxter relation \cite{yang, baxter}:
\par
$\mathcal{R}_{\,\,\,\,\,\,a_{2}b_{2}}^{a_{1}b_{1}}\mathcal{R}
_{\,\,\,\,\,\,a_{3}c_{2}}^{a_{2}c_{1}}\mathcal{R}_{\,\,\,\,\,
\,b_{3}c_{3}}^{b_{2}c_{2}}=\mathcal{R}_{\,\,\,\,\,\,b_{2}c_{2}}^{b_{1}c_{1}} 
\mathcal{R}_{\,\,\,\,\,\,a_{2}c_{3}}^{a_{1}c_{2}}\mathcal{R}
_{\,\,\,\,\,\,a_{3}b_{3}}^{a_{2}b_{2}}$.
\par
If $\left( \mathcal{A},\mathcal{R}\right) $ is a quasitriangular Hopf
algebra, then $\mathcal{R}$ satisfies the Yang-Baxter equation.
\par
The Yang Baxter equation and the noncommutativity of the elements $%
M_{\,\,\,\,m}^{n}$ \ lie at the foundation of the method of commuting
transfer-matrices in classical statistical mechanics \cite{baxter} and
factorizable scattering theory \cite{yang,zamolodchikov}, the quantum theory
of magnets \cite{bethe} and the inverse scattering method for solving
nonlinear equations of evolutions \cite{gardner}.}

\begin{equation}
R_{\,\,\,\,\,kl}^{ij}=\left( 
\begin{array}{cccc}
1 & 0 & 0 & 0 \\ 
0 & 0 & q & 0 \\ 
0 & q & 1-q^{2} & 0 \\ 
0 & 0 & 0 & 1%
\end{array}
\right) .
\end{equation}

Let us give the example of $U_{q}\left( 2\right) $ obtained by requiring
that the unitary condition hold for this $2\times 2$ \ quantum matrix:

\begin{equation}
M_{\,\,\,\,\,\,\,m}^{n\dagger }=M_{\,\,\,\,\,\,\,m}^{n\,\, -1}.
\end{equation}

The $2\times 2$ matrix belonging to $U_{q}\left( 2\right) $ preserves the \
nondegenerate bilinear form \cite{dubois} $C_{nm}$

\begin{equation}
C_{nm}M_{\,\,\,k}^{n}M_{\,\,\,\,\,l}^{m}=D_{q}C_{kl},\quad
C^{nm}M_{\,\,\,n}^{k}M_{\,\,\,m}^{l}=D_{q}C^{kl},\quad C_{kn}C^{nl}=\delta
_{k}^{l},
\end{equation}

\begin{equation}
C_{nm}=\left( 
\begin{array}{cc}
0 & -q^{-1/2} \\ 
q^{1/2} & 0%
\end{array}
\right) ,\quad \,C^{nm}=\left( 
\begin{array}{cc}
0 & q^{-1/2} \\ 
-q^{1/2} & 0%
\end{array}
\right) .\,
\end{equation}

\bigskip

The algebra $Fun\left( U_{q}\left( 2\right) \right) $ is freely generated by
the associative unital $C^{\ast }$-algebra. $Fun\left( U_{q}\left( 2\right)
\right) $ is a Hopf algebra with comultiplication\ $\Delta $, counit $%
\epsilon $ and antipode $S$ which are given by:

-comultiplication (also called coproduct)

\begin{equation}
\Delta \left( M_{\,\,\,m}^{n}\right) =M_{\,\,\,k}^{n}\otimes M_{\,\,\,m}^{k}.
\end{equation}

This coproduct $\Delta $ on $Fun\left( U_{q}\left( 2\right) \right) $ is
directly related, for $q=1$ (the nondeformed case), to the pullback induced
by left multiplication of the group on itself.

-co-unit $\epsilon $

\begin{equation}
\varepsilon \left( M_{\,\,\,m}^{n}\right) =\delta _{m}^{n}
\end{equation}

-antipode $S$ (coinverse)

\begin{equation}
S\left( M_{\,\,\,k}^{n}\right) M_{\,\,\,m}^{k}=M_{\,\,\,k}^{n}S\left(
M_{\,\,\,m}^{k}\right) =\delta _{\,m}^{n}
\end{equation}

\begin{equation}
S\left( M_{\,\,\,m}^{n}\right) =\frac{1}{D_{q}}C^{nk}M_{\,\,k}^{l}C_{lm}.
\end{equation}

With the nondegenerate form $C$ the $R$ matrix has the form

\begin{eqnarray}
R_{\,\,\,\,\,\,\,\,\,\,\,\,\,\,\,\,kl}^{+nm} &=&R_{\quad \,kl}^{nm}=\delta
_{k}^{n}\delta _{l}^{m}+qC^{nm}C_{kl}, \\
R_{\,\,\,\,\,\,\,\,\,\,\,\,\,\,\,kl}^{-nm}
&=&R_{\,\,\,\,\,\,\,\,\,\,\,\,\,\,\,\,kl}^{-1nm}=\delta _{k}^{n}\delta
_{l}^{m}+q^{-1}C^{nm}C_{kl}.
\end{eqnarray}

The $R$ matrices satisfy the Hecke relations

\begin{equation}
R^{\pm 2}=\left( 1-q^{\pm 2}\right) R^{\pm }+q^{\pm 2}\mathbf{1}
\end{equation}

and the relations

\begin{equation}
C_{nm}R_{\qquad kc}^{\pm an}R_{\qquad lb}^{\pm cm}=q^{\pm 1}\delta
_{b}^{a}C_{kl}.
\end{equation}

\bigskip

\section{Review of Woronowicz's Bicovariant Differential Calculus}

In this section we give a short review of the bicovariant differential
calculus on quantum groups as developed by Woronowicz \cite{woronowicz}. 
\newline

\textit{Definition:} \newline
A \textbf{first-order differential calculus over an algebra} $\mathcal{A}$
is a pair $\left( \Gamma ,d\right) $ such that:

\begin{description}
\item (1) $\Gamma $ is an $\mathcal{A}$-bimodule, i.e. \ $\left( a\omega
\right) b=a\left( \omega b\right) $
\end{description}

for all $a,b\in \mathcal{A}$, $\omega \in \Gamma $, where the left and right
actions which make $\Gamma $, respectively a left $\mathcal{A}$-module and a
right $\mathcal{A}$-module are written multiplicatively;

\begin{description}
\item (2) $d$ is a linear map, $d:\mathcal{A}\rightarrow \Gamma $;

\item (3) for any $a,b\in \mathcal{A}$, the Leibniz rule is satisfied, i.e. 
\begin{equation}
d\left( ab\right) =d\left( a\right) b+ad\left( b\right)
\end{equation}

\item (4) the bimodule $\Gamma $, or ``space of one-forms'', is spanned by
elements of the form $adb$, $a,b\in \mathcal{A}$.
\end{description}

\bigskip

\textit{Definition:} \newline
A \textbf{bicovariant bimodule} over a Hopf algebra $\mathcal{A}$ is a
triple $\left( \Gamma ,\Delta _{L},\Delta _{R}\right) $ such that :

\begin{description}
\item (1) $\Gamma $ is an $\mathcal{A}$-bimodule;

\item (2) $\Gamma $ is an $\mathcal{A}$-bicomodule with left and right
coactions $\Delta _{L}$ and $\Delta _{R}$ respectively, i.e.
\end{description}

\begin{equation}
\left( id\otimes \Delta _{L}\right) \circ \Delta _{L}=\left( \Delta \otimes
id\right) \circ \Delta _{L}\qquad \left( \epsilon \otimes id\right) \circ
\Delta _{L}=id
\end{equation}

making $\Gamma $ a left $\mathcal{A}$-comodule,

\begin{equation}
\left( \Delta _{R}\otimes id\right) \circ \Delta _{R}=\left( id\otimes
\Delta \right) \circ \Delta _{R}\qquad \left( id\otimes \epsilon \right)
\circ \Delta _{R}=id
\end{equation}

making $\Gamma $ a right $\mathcal{A}$-comodule, and

\begin{equation}
\left( id\otimes \Delta _{R}\right) \circ \Delta _{L}=\left( \Delta
_{L}\otimes id\right) \circ \Delta _{R}
\end{equation}

which is a the $\mathcal{A}$-bicomodule property;

\begin{description}
\item (3) the coactions \footnote{%
The left (resp. right) coactions are pullbacks for one forms induced by left
(resp. right) multplication of the group by itself.}, $\Delta _{L}$ and $%
\Delta _{R}$ are bimodule maps, i.e.
\end{description}

\begin{eqnarray}
\Delta _{L}\left( a\omega b\right) &=&\Delta \left( a\right) \Delta
_{L}\left( \omega \right) \Delta \left( b\right)  \notag \\
\Delta _{R}\left( a\omega b\right) &=&\Delta \left( a\right) \Delta
_{R}\left( \omega \right) \Delta \left( b\right)
\end{eqnarray}

\bigskip

\textit{Remark: } \newline
The Sweedler notation for coproducts in the Hopf algebra $\mathcal{A}$ is
taken to be $\Delta \left( a\right) =a_{\left( 1\right) }\otimes a_{\left(
2\right) }$ for all $a\in \mathcal{A}$ and is extended to the coactions as $%
\Delta _{L}\left( \omega \right) =\omega _{\left( \mathcal{\ A}\right)
}\otimes \omega _{\left( \Gamma \right) }$ and $\Delta _{R}\left( \omega
\right) =\omega _{\left( \Gamma \right) }\otimes \omega _{\left( \mathcal{A}%
\right) }$.

\bigskip

\textit{Definition: } \newline
A \textbf{first-order bicovariant differential calculus over a Hopf algebra} 
$\mathcal{A}$ is a quadruple $\ \left( \Gamma ,d,\Delta _{L},\Delta
_{R}\right) $ such that:

\begin{description}
\item (1) $\left( \Gamma ,d\right) $ is a first-order differential calculus
over $\mathcal{A}$;

\item (2) $\left( \Gamma ,\Delta _{L},\Delta _{R}\right) $ is a bicovariant
bimodule over $\mathcal{A}$;

\item (3) $d$ is both a left and a right comodule map, i.e.
\end{description}

\begin{eqnarray}
\left( id\otimes d\right) \circ \Delta \left( a\right) &=&\Delta _{L}\left(
da\right)  \notag \\
\left( d\otimes id\right) \circ \Delta \left( a\right) &=&\Delta _{R}\left(
da\right)
\end{eqnarray}

for all $a\in \mathcal{A}$.

\section{$SO_{q}\left( 6\right) $ Bicovariant Differential Calculus}

In this section, we construct the left invariant vector fields, the quantum
trace and the quantum Killing metric which are the important ingredients in
the construction of quantum gauge theories. \newline
Let us first consider the bicovariant bimodule $\Gamma $ over the quantum
group $SO_{q}\left( 6\right) $. This quantum group is the symmetry group of
the $q$-deformed AdS/CFT correspondence (to be constructed). The
corresponding braiding matrix is given by:

\begin{eqnarray}
\hat{R} &=&q\sum_{\overset{i=-3}{i\neq 0}}^{3}\delta _{i}^{i}\otimes \delta
_{i}^{i}+\sum_{\overset{i,j=-3}{i\neq j,-j}}^{3}\delta _{i}^{i}\otimes
\delta _{j}^{j}+q^{-1}\sum_{\overset{i=-3}{i\neq 0}}^{3}\delta
_{-i}^{-i}\otimes \delta _{i}^{i}  \notag \\
&+&k\sum_{\overset{i,j=-3}{i>j}}^{3}\delta _{j}^{i}\otimes \delta
_{i}^{j}-k\sum_{\overset{i,j=-3}{i>j}}^{3}q^{\rho _{i}-\rho _{j}}\delta
_{j}^{i}\otimes \delta _{-j}^{-i}
\end{eqnarray}%
where we have used the notations: 
\begin{eqnarray}
k &\equiv &q-q^{-1}  \notag \\
\rho _{i} &=&(2,1,0,0,-1,-2).
\end{eqnarray}%
The matrix elements $\hat{R}$ vanish unless the indices satisfy the
following conditions: 
\begin{eqnarray}
\mathrm{either} &&\,\,\,i\neq -j\quad \mathrm{and}\quad k=i,\,\,l=j,\quad 
\mathrm{or}\quad l=i,k=j  \notag \\
\mathrm{or} &&\,\,\,i=-j\quad \mathrm{and}\quad k=-l.
\end{eqnarray}

The matrix $\hat{R}$ enters in local representations of the
Birman-Wenzel-Murakami algebra \cite{birman}. $\hat{R}$ admits a projector
decomposition \cite{faddeev1}: 
\begin{equation}
\hat{R}=qP_{S}-q^{-1}P_{a}+q^{-5}P_{t},
\end{equation}
where $P_{S}$, $P_{a}$, $P_{t}$ are the projections operators onto three
eigenspaces of $\hat{R}$ with dimensions respectively 20, 15, 1: they
project the tensor product $x\otimes x$ of the fundamental corepresentation $%
x$ of $SO_{q}\left( 6\right) $ into the corresponding irreducible
corepresentations:

\begin{eqnarray}
P_{S}&=&\frac{1}{q+q^{-1}}\left[ \hat{R}+q^{-1}I-\left( q^{-1}+q^{-5}\right)
P_{t}\right] ,  \notag \\
P_{a}&=&\frac{1}{q+q^{-1}}\left[ -\hat{R}+qI-\left( q+q^{-5}\right) P_{t}%
\right],  \notag \\
\left( P_{t}\right) ^{ab} _{cd} &=& \left( C_{ef}C^{ef}\right) ^{-1}
C_{cd}C^{ab} .
\end{eqnarray}

The $\hat{R}$ matrices satisfy the Yang-Baxter equation.

\begin{equation}
\hat{R}^{ij} _{\,\,\,\,\,\,pq}\hat{R}^{qk} _{\,\,\,\,\,\,mn}\hat{R}^{pm}
_{\,\,\,\,\,\,\,rl}=\hat{R}^{jk} _{\,\,\,\,\,\,mp}\hat{R}^{im}
_{\,\,\,\,\,\,rq}\hat{R}^{qp} _{\,\,\,\,\,\,ln}
\end{equation}
and the relations: 
\begin{eqnarray}
C_{qm}\hat{R}^{\pm qp} _{\,\,\,\,\,\,\,\,\,\,ji}\hat{R}^{\pm ml}
_{\,\,\,\,\,\,\,\,\,\,pk}&=&\delta ^{l} _{j}C_{ik}  \notag \\
C^{qm}\hat{R}^{\pm ip} _{\,\,\,\,\,\,\,\,\,\,jq}\hat{R}^{\pm kl}
_{\,\,\,\,\,\,\,\,\,\,pm}&=&\delta ^{l} _{j}C^{ik}.
\end{eqnarray}
The noncommutativity of the elements $M^{i} _{\,\,\,j}$ is expressed as 
\begin{equation}
\hat{R}^{ij} _{\,\,\,\,\,\,pq}M^{p} _{\,\,\,l}M^{q} _{\,\,\,n}=M^{i}
_{\,\,\,q}M^{j} _{\,\,\,p}\hat{R}^{qp} _{\,\,\,\,\,\,ln}.
\end{equation}
The generators $M^{i} _{\,\,\,j}$ satisfy the orthogonality condition 
\begin{equation}
C_{ij}M^{i} _{\,\,\,q}M^{j} _{\,\,\,p}=C_{qp}.
\end{equation}
The nondegenerate bilinear form $C^{nm}$ is given by 
\begin{equation}
C^{nm}=\left( 
\begin{array}{cccccc}
0 & 0 & 0 & 0 & 0 & q^{-2} \\ 
0 & 0 & 0 & 0 & q^{-1} & 0 \\ 
0 & 0 & 0 & 1 & 0 & 0 \\ 
0 & 0 & 1 & 0 & 0 & 0 \\ 
0 & q & 0 & 0 & 0 & 0 \\ 
q^{2} & 0 & 0 & 0 & 0 & 0%
\end{array}
\right) .
\end{equation}
Now, let us consider the fundamental bimodule over $Fun\left( SO_{q}\left( 6
\right) \right) $ generated by left invariant basis $\theta ^{a}$, $%
a=1,...,6 $. The right coaction is defined as 
\begin{equation}
\Delta _{R}\left( \theta ^{a}\right)=\theta ^{b}\otimes S\left( M^{a}
_{\,\,\,b}\right) .
\end{equation}
We can also define the conjugate $\left( \theta ^{*a}\right) =\left( \theta
^{a}\right) ^{\ast }\equiv \bar{\theta }_{a}$. The right coaction acts on
this basis as 
\begin{equation}
\Delta _{R}\left( \bar{\theta _{a}}\right) = \bar{\theta }_{b}\otimes M^{b}
_{\,\,\,a} .
\end{equation}

This equation is easily obtained from Equ. (46) by the antilinear $*$
involution using the relation $\left( \Delta _{R}\left( \theta ^{a}\right)
\right) ^{*}=\Delta _{R}\left( \theta ^{a}\right) ^{*}, \quad \left(M^{a}
_{\,\,\,b}\right) ^{*}\equiv M^{\dagger b} _{\,\,\,\,\,a}=S\left( M^{b}
_{\,\,\,a}\right) $. There exist linear functionals $f^{a} _{\,\,\,b}$ and $%
\bar{f}^{a} _{\,\,\,b}:Fun\left( SO_{q}\left( 6\right) \right) \rightarrow 
\mathbb{C}$ for the left basis $\theta ^{a}$ and $\bar{\theta }_{a}$ such
that 
\begin{eqnarray}
\theta ^{a}M^{n} _{\,\,\,m}&=&\left( f^{a} _{\,\,\,b}\star M^{n}
_{\,\,\,m}\right) \theta ^{b}=\left( id\otimes f^{a} _{\,\,\,b}\right)
\Delta \left( M^{n} _{\,\,\,m}\right) \theta ^{b}  \notag \\
&=& f^{a} _{\,\,\,b}\left( M^{k} _{\,\,\,m}\right) M^{n} _{\,\,\,k}\theta
^{b} ,
\end{eqnarray}
\begin{eqnarray}
M^{n} _{\,\,\,m}\theta ^{a}&=& \theta ^{b}\left( f^{a} _{\,\,\,b}\circ
S\star M^{n} _{\,\,\,m}\right)= \theta ^{b}\left( id\otimes f^{a}
_{\,\,\,b}\circ S\right) \Delta \left( M^{n} _{\,\,\,m}\right)  \notag \\
&=& f^{a} _{\,\,\,b}\left( S\left( M^{k} _{\,\,\,m}\right) \right) \theta
^{b}M^{n} _{\,\,\,k} ,
\end{eqnarray}
\begin{eqnarray}
\bar{\theta }_{a}M^{n} _{\,\,\,m}&=&\left( \bar{f}^{b} _{\,\,\,a}\star M^{n}
_{\,\,\,m}\right) \bar{\theta }_{b}=\left( id\otimes \bar{f}^{b}
_{\,\,\,a}\right) \Delta \left( M^{n} _{\,\,\,m}\right) \bar{\theta }_{b} 
\notag \\
&=& \bar{f}^{b} _{\,\,\,a}\left( M^{k} _{\,\,\,m}\right) M^{n} _{\,\,\,k}%
\bar{\theta }_{b} ,
\end{eqnarray}
\begin{eqnarray}
M^{n} _{\,\,\,m}\bar{\theta }_{a}&=&\bar{\theta }_{b} \left( \bar{f}^{b}
_{\,\,\,a}\circ S\star M^{n} _{\,\,\,m}\right) =\bar{\theta }_{b}\left(
id\otimes \bar{f}^{b} _{\,\,\,a}\circ S\right) \Delta \left( M^{n}
_{\,\,\,m}\right)  \notag \\
&=& \bar{f}^{b} _{\,\,\,a}\left( S\left( M^{k} _{\,\,\,m}\right) \right) 
\bar{\theta }_{b}M^{n} _{\,\,\,k} .
\end{eqnarray}

The orthogonality condition Equ. (44) must be consistent with the bimodule
structure. This implies that 
\begin{equation}
f^{a} _{\,\,\,b}\left( C_{ij}M^{i} _{\,\,\,q}M^{j} _{\,\,\,p}\right)
=C_{ij}f^{a} _{\,\,\,c}\left( M^{i} _{\,\,\,q}\right) f^{c} _{\,\,\,b}\left(
M^{j} _{\,\,\,p}\right) =\delta ^{a}_{b}C_{qp} .
\end{equation}

Comparing with Equ. (42) we get two solutions

\begin{eqnarray}
f^{a} _{+\,\,\,b}\left( M^{n} _{\,\,\,m}\right) &=& \hat{R}^{an}
_{\,\,\,\,\,\,mb} \\
f^{a} _{-\,\,\,b}\left( M^{n} _{\,\,\,m}\right) &=& \hat{R}^{-an}
_{\,\,\,\,\,\,\,\,\,\,mb} .
\end{eqnarray}

Applying the $*$-operation on both sides of Equ. (48) and substituting $%
\left( M^{a} _{\,\,\,b}\right) ^{*}\equiv M^{\dagger b}
_{\,\,\,\,\,\,a}=S\left( M^{b} _{\,\,\,a}\right) $ we get

\begin{equation}
\bar{f}^{a} _{\,\,\,\pm b}\left( S\left( M^{n} _{\,\,\,m}\right) \right) =
f^{a} _{\,\,\,\mp b}\left( M^{n} _{\,\,\,m}\right) = \hat{R}^{\mp an}
_{\,\,\,\,\,\,mb} .
\end{equation}

The representation with upper index of $\bar{\theta }_{a}$ is defined by the
bilinear form $C$: 
\begin{equation}
\bar{\theta }^{b} _{\pm}=\bar{\theta }_{\pm a}C^{ab} .
\end{equation}

Then the right coaction is 
\begin{equation}
\Delta _{R}\left( \bar{\theta }^{b} _{\pm }\right) = \bar{\theta }^{a} _{\pm
}\otimes C_{ad}M^{d} _{\,\,\,e}C^{eb}=\bar{\theta }^{a} _{\pm }\otimes
S^{-1}\left( M^{b} _{\,\,\,a}\right) ,
\end{equation}
where the inverse of the antipode $S^{-1}$ satisfies 
\begin{equation}
S^{-1}S\left( M^{a} _{\,\,\,b}\right) =M^{a} _{\,\,\,b} .
\end{equation}
The functionals $\tilde{f} ^{a} _{\,\,\,\pm b}$ corresponding to the basis $%
\bar{\theta }^{b} _{\pm}$ are given by 
\begin{equation}
\tilde{f} ^{a} _{\,\,\,\pm b}=C_{bd}\bar{f}^{d} _{\,\,\,e}C^{ea} .
\end{equation}
The transformation of the adjoint representation for the quantum group acts
on the generators $M^{a} _{\,\,\,b}$ as the right adjoint coaction\footnote{%
Example: Let $\mathcal{A}=\mathcal{O}\left( G\right) $ a coordinate Hopf
algebra, then 
\begin{eqnarray*}
\forall g,h \in G, Ad_{R}\left( M^{i} _{\,\,\,j}\right) \left( g,h\right)
&=& \sum M^{k} _{\,\,\,l}\left( g\right) \left( S\left( M^{i}
_{\,\,\,k}\right) M^{l} _{\,\,\,j}\right) \left( h\right) = \sum g^{k}
_{\,\,\,l}\left( h^{-1}\right) ^{i} _{\,\,\,k}h^{l} _{\,\,\,j} \\
&=& \left( h^{-1}gh\right) ^{i} _{\,\,\,j}=M^{i} _{\,\,\,j}\left(
h^{-1}gh\right) .
\end{eqnarray*}%
} $Ad_{R}$ : 
\begin{equation}
Ad_{R}\left( M^{n} _{\,\,\,\,m}\right) =M^{l} _{\,\,\,k}\otimes S\left(
M^{n} _{\,\,\,l}\right) M^{k} _{\,\,\,m} .
\end{equation}
A bicovariant bimodule which includes the adjoint transformation $\Gamma
_{Ad}$ is obtained by taking the tensor product $\theta ^{n}\otimes \bar{%
\theta } _{m}\equiv \theta ^{n} _{\,\,\,m}$ of two fundamental modules. 
\newline
The right coaction $\Delta _{R}$ on the basis $\theta ^{n} _{\,\,\,m}$ is

\begin{equation}
\Delta _{R}\left( \theta ^{n} _{\,\,\,m}\right) =\theta ^{l}
_{\,\,\,k}\otimes S\left( M^{n} _{\,\,\,l}\right) M^{k} _{\,\,\,m} .
\end{equation}

With the requirement that the $\ast $-operation be a module antiautomorphism 
$\left( \Gamma _{Ad}\right) ^{\ast }=\Gamma _{Ad}$ we can find two different
types of left invariant basis containing the adjoint representation. We take 
$\theta _{+\,\,\,\,m}^{n}=\theta _{+}^{n}\bar{\theta}_{+m}$.

We introduce the left invariant basis $\theta ^{ab}$ with upper indices:

\begin{equation}
\theta ^{ab}=\theta ^{a} _{\,\,\,c}C^{cb}.
\end{equation}

In this basis the right coaction is given by

\begin{equation}
\Delta _{R}\left( \theta ^{ab}\right) = \theta ^{cd}\otimes S\left( M^{a}
_{\,\,\,c}\right) S^{-1}\left( M^{b} _{\,\,\,d}\right) .
\end{equation}

The relation between the left and right multiplication for this basis is

\begin{eqnarray}
\theta ^{ab}M^{n} _{\,\,\,m}&=&\left( \theta ^{a}\otimes \bar{\theta }%
^{b}\right)M^{n} _{\,\,\,m}=\theta ^{a}\otimes \left( \tilde{f}^{b}
_{\,\,\,d}\star M^{n} _{\,\,\,m}\right) \bar{\theta }^{d}  \notag \\
&=&\left( f^{a} _{\,\,\,c}\star \left( \tilde{f}^{b} _{\,\,\,d}\star M^{n}
_{\,\,\,m}\right) \right) \left( \theta ^{c}\otimes \bar{\theta }^{d} \right)
\notag \\
&=&\left( f^{ab} _{Ad\,\,\,\,\,\,cd}\star M^{n} _{\,\,\,m}\right) \theta
^{cd}=f^{ab} _{Ad\,\,\,\,\,\,cd}\left( M^{k} _{\,\,\,m}\right) M^{n}
_{\,\,\,k} \theta ^{cd}
\end{eqnarray}
where 
\begin{eqnarray}
f^{ab} _{Ad\,\,\,\,\,\,cd}\left( M^{n} _{\,\,\,m}\right) &=&\left( f^{a}
_{\,\,\,c}\star \tilde{f}^{b} _{\,\,\,d}\right) \left( M^{n}
_{\,\,\,m}\right) =\left( f^{a} _{\,\,\,c}\otimes \tilde{f}^{b}
_{\,\,\,d}\right) \Delta \left( M^{n} _{\,\,\,m}\right)  \notag \\
&=&f^{a} _{\,\,\,c}\left( M^{n} _{\,\,\,k}\right) \tilde{f}^{b}
_{\,\,\,d}\left( M^{k} _{\,\,\,m}\right) = \hat{R}^{an}
_{\,\,\,\,\,\,kc}C_{de}\bar{f}^{e} _{\,\,\,f}\left( M^{k} _{\,\,\,m}\right)
C^{fb}  \notag \\
&=&\hat{R}^{an} _{\,\,\,\,\,\,kc}\hat{R}^{-bk} _{\,\,\,\,\,\,\,\,md} .
\end{eqnarray}
The exterior derivative $d$ is defined as 
\begin{eqnarray}
dM^{n} _{\,\,\,m}&=&\frac{1}{\mathcal{N}}\left[ X, M^{n} _{\,\,\,m}\right]
_{-}=\left( \chi _{ab}\star M^{n} _{\,\,\,m}\right) \theta ^{ab}  \notag \\
&=&\left( id \otimes \chi _{ab}\right) \Delta \left( M^{n} _{\,\,\,m}\right)
\theta ^{ab}=\chi _{ab}\left( M^{k} _{\,\,\,m}\right) M^{n} _{\,\,\,k}\theta
^{ab}
\end{eqnarray}
where $X=C_{ab}\theta ^{ab}$ is the singlet representation of $\theta ^{ab}$
and is both left and right co-invariant, $\mathcal{N}\in \mathbb{C}$ is a
normalization constant which we take purely imaginary $\mathcal{N}^{*}=-%
\mathcal{N}$ and $\chi _{ab}$ are the quantum analogue of left-invariant
vector fields given by: 
\begin{eqnarray}
\chi _{ab}\left( M^{n} _{\,\,\,m}\right) &=&\frac{1}{\mathcal{N}}\left(
C_{cd}f^{cd} _{Ad\,\,\,ab}\left( M^{n} _{\,\,\,m}\right) - \delta ^{n}
_{m}C_{ab}\right)  \notag \\
&=&\frac{1}{\mathcal{N}}\left( C_{cd}\hat{R}^{cn} _{\,\,\,\,\,ka}\hat{R}%
^{-dk} _{\,\,\,\,\,\,\,mb}- \delta ^{n} _{m}C_{ab}\right) .
\end{eqnarray}

Higher order differential calculus is built from the first order
differential calculus by using the tensor product $\Gamma _{Ad}\otimes
\Gamma _{Ad}\otimes ...\otimes \Gamma _{Ad}$. The basic operation is the
bicovariant bimodule automorphism $\Lambda :\Gamma ^{\otimes 2}\rightarrow
\Gamma ^{\otimes 2}$, defined as 
\begin{equation}
\Lambda \left( \theta ^{ab}\otimes \omega ^{cd}\right) =\omega ^{cd}\otimes
\theta ^{ab},
\end{equation}%
where $\omega ^{ab}$ is the right invariant basis defined by 
\begin{equation}
\omega ^{ab}=M_{\,\,\,c}^{a}M_{\,\,\,d}^{b}\theta ^{cd}.
\end{equation}%
Let $I$ represents a set of indices $I=\left( a,b\right) $, we write Equ.
(69) as

\begin{equation}
\omega ^{I}=T^{I} _{\,\,\,J}\theta ^{J},
\end{equation}

where $T^{I} _{\,\,\,J}=T^{ab} _{\,\,\,\,\,cd}=M^{a} _{\,\,\,c}M^{b}
_{\,\,\,d}$. \newline
Using Equ. (69) and the definition of a bicovariant bimodule automorphism
i.e. $\Lambda \left( a \tau b\right) = a \Lambda \left( \tau \right) b,
\quad \forall a,b\in \Gamma ^{\otimes 2} _{Ad}:$

\begin{eqnarray}
\Lambda \left( \theta ^{I}\otimes T^{J} _{\,\,\,K}\theta ^{K}\right)
&=&T^{J} _{\,\,\,K}\theta ^{K} \otimes \theta ^{I}  \notag \\
&=&f^{I} _{Ad\,\,\,\,\,N}\left( T^{L} _{\,\,\,K}\right) T^{J}
_{\,\,\,L}\Lambda \left( \theta ^{N} \otimes \theta ^{K}\right) ,
\end{eqnarray}
which gives 
\begin{eqnarray}
\Lambda \left( \theta ^{M} \otimes \theta ^{L}\right) &=&f^{M}
_{Ad\,\,\,\,\,J}\left( S\left( T^{L} _{\,\,\,K}\right) \right) \left( \theta
^{K}\otimes \theta ^{J}\right)  \notag \\
&=&\Lambda ^{ML} _{\,\,\,\,\,\,\,\,\,\,\,KJ}\left( \theta ^{K}\otimes \theta
^{J}\right) .
\end{eqnarray}
Therefore, the matrix representation of $\Lambda $ on the basis $\theta
^{ab}\otimes \theta ^{cd}$ is 
\begin{equation}
\Lambda ^{ML} _{\,\,\,\,\,\,\,\,\,\,\,KJ}=f^{M} _{Ad\,\,\,\,\,J}\left(
S\left( T^{L} _{\,\,\,K}\right) \right) ,
\end{equation}
leading to 
\begin{equation}
\Lambda ^{ijkl} _{\,\,\,\,\,\,\,\,\,\,\,ghef}=f^{ij} _{Ad \,\,\,\,ef}\left(
S\left( M^{k} _{\,\,\,g}M^{l} _{\,\,\,h}\right) \right) .
\end{equation}
The action of the exterior derivative $d$ on $\mathcal{A}$ can be
generalized on $p$-forms as in the usual differential calculus: 
\begin{equation}
d:\Gamma ^{\wedge p} _{Ad}\rightarrow \Gamma ^{\wedge p+1} ,
\end{equation}
and is defined by $\forall \Omega \in \Gamma ^{\wedge p} _{Ad} :$ 
\begin{equation}
d \Omega \equiv \frac{1}{\mathcal{N}}\left[ X,\Omega \right] _{\pm}=\frac{1}{%
\mathcal{N}}\left( X \wedge \Omega - \left( -1\right) ^{p} \Omega \wedge
X\right) .
\end{equation}
The external product is given by 
\begin{equation}
\theta ^{ab}\wedge \theta ^{cd}=\left( \delta ^{a} _{e}\delta ^{b}
_{f}\delta ^{c} _{g}\delta ^{d} _{h} - \Lambda ^{abcd}
_{\,\,\,\,\,\,\,\,\,\,efgh}\right) \left( \theta ^{ef}\otimes \theta
^{gh}\right) .
\end{equation}
The two-form has been defined as 
\begin{equation}
\Gamma ^{\wedge 2} _{Ad}=\Gamma ^{\otimes 2} _{Ad} / \left[ \mathrm{Ker}
\left( \Lambda -1\right) \oplus \mathrm{Ker} \left( \Lambda -q^{-6}I\right) %
\right] .
\end{equation}
This equation can be expressed in terms of the projectors and gives the
Cartan-Maurer equations. \newline
The quantum commutators of the quantum Lie algebra generators $\chi _{ab}$
are defined as 
\begin{equation}
\left[ \chi _{ab},\chi _{cd}\right] \left( M^{i} _{\,\,\,j}\right) = \left(
1-\Lambda \right) ^{efgh} _{\,\,\,\,\,\,\,\,\,\,\,abcd}\left( \chi _{ef}
\star \chi _{gh} \right)
\end{equation}
and can be written as 
\begin{eqnarray}
\left[ \chi _{ab},\chi _{cd}\right] \left( M^{i} _{\,\,\,j}\right) &=&\left(
\chi _{ab}\otimes \chi _{cd}\right) Ad_{R}\left( M^{i}_{\,\,\,j} \right) 
\notag \\
&=& \chi _{ab}\left( M^{l} _{\,\,\,n}\right) \otimes \chi _{cd}\left(
S\left( M^{i} _{\,\,\,l}\right) M^{n} _{\,\,\,j}\right)  \notag \\
&=& C_{abcd} ^{\,\,\,\,\,\,\,\,\,\,\,\,ef}\chi _{ef}\left( M^{i}
_{\,\,\,j}\right) ,
\end{eqnarray}
where $C_{abcd} ^{\,\,\,\,\,\,\,\,\,\,\,\,ef}$ are the quantum structure
constants. \newline

To construct a quantum gauge invariant Lagrangian, we need a well defined
quantum trace. We require that this trace is invariant under the right
adjoint coaction:

\begin{equation}
Tr\left( M^{i} _{\,\,\,j}\right) =Tr\left( Ad_{R}\left( M^{i}
_{\,\,\,j}\right) \right) = Tr \left( M^{l} _{\,\,\,n}\otimes S\left( M^{i}
_{\,\,\,l}\right) M^{n} _{\,\,\,j}\right) .
\end{equation}

This equation is fulfilled if one defines the quantum trace as 
\begin{equation}
Tr\left( M^{i} _{\,\,\,j}\right) =- C_{nk}M^{n} _{\,\,\,m}C^{mk} .
\end{equation}
The quantum trace allows us to introduce the quantum Killing metric as in
the usual undeformed case ($q=1$) 
\begin{equation}
g_{abcd}=Tr\left( \chi _{ab}\left( M^{n} _{\,\,\,k}\right) \chi _{cd} \left(
M^{k} _{\,\,\,m}\right) \right) .
\end{equation}

\bigskip

Before we close this section, we note that the same construction can be done
using the adjoint representation $M^{\,\,\,\,a} _{b}$ of the quantum group
(see the paper of Aschieri and Castellani \cite{books}). This allows us to
find the relation between the Cartan-Maurer forms $\tilde{\theta }^{i}
_{\,\,\,j}$ and the forms obtained by taking the tensor product $\theta
^{ab}=\theta ^{a}\bar{\theta } ^{b}$ .

Let $\tilde{\theta ^{a}}$ be a left invariant basis of $_{\mathrm{inv}}
\Gamma $, the linear subspace of all left-invariant elements of $\Gamma $
i.e. $\Delta _{L}\left( \tilde{\theta ^{a}}\right) =I\otimes \tilde{\theta
^{a}}$. In the case $q=1$ the left coaction $\Delta _{L}$ coincides with the
pullback for 1-forms induced by the left multiplication of the group on
itself. There exists an adjoint representation $M_{b} ^{\,\,\,a}$ of the
quantum group, defined by the right coaction $\Delta _{R}$ on the
left-invariant $\tilde{\theta ^{a}}$ : 
\begin{equation}
\Delta _{R}\left( \tilde{\theta ^{a}}\right) = \tilde{\theta ^{b}}\otimes
M_{b} ^{\,\,\,a}, \quad \quad M_{b} ^{\,\,\,a}\in \mathcal{A}
\end{equation}
where $\mathcal{A}$ is an associative unital $\mathbb{C} $-algebra. In the
classical case, $M_{b} ^{\,\,\,a}$ is indeed the adjoint representation of
the group $SO\left( 6 \right) $. We recall that in this limit the
left-invariant 1-forms $\tilde{\theta ^{a}}$ can be constructed as 
\begin{equation}
\tilde{\theta ^{a}}\left( y\right) T_{a}=\left(y^{-1}dy\right) ^{a} \quad
\quad y \in SO\left( 6 \right) .
\end{equation}
Under the right multiplication by a (constant) element $x\in SO\left(
6\right) :y\rightarrow x$ we have\footnote{%
We recall that the $q=1$ adjoint representation is defined as: $%
x^{-1}T_{b}x\equiv M_{b} ^{\,\,\,a}\left( x\right)T_{a}$ where the
infinitesimal operators carry the adjoint representation of the Lie group $%
SO\left( 6\right) $.}, 
\begin{eqnarray}
\tilde{\theta ^{a}}\left( yx\right) T_{a}&=&\left[ x^{-1}y^{-1}d\left(
yx\right) \right] ^{a}T_{a}=\left[ x^{-1}\left( y^{-1}dy\right) x\right]
^{a}T_{a}  \notag \\
&=&\left[ x^{-1}T_{b}x\right] ^{a}\left( y^{-1}dy\right) ^{b}T{_a}= M_{b}
^{\,\,\,a}\left( x\right) \tilde{\theta ^{b}}\left( y\right) T_{a} ,
\end{eqnarray}
so that 
\begin{equation}
\tilde{\theta ^{a}}\left(yx\right)=\tilde{\theta ^{b}}\left(y\right) M_{b}
^{\,\,\,a}\left( x\right)
\end{equation}
or

\begin{equation}
R^{\star }\tilde{\theta ^{a}}\left(y,x\right) = \tilde{\theta ^{b}} M_{b}
^{\,\,\,a}\left(y,x\right) ,
\end{equation}

which reproduces Equ.(84) for $q=1$ ($R^{\star }$ is the analogue of the
right coaction $\Delta _{R}$ in this limit and is defined via the pullback $%
R^{\star } _{x}$ on functions or 1-forms induced by right multiplication of
the group $SO\left( 6\right) $ on it self).

To obtain the adjoint representation $M_{b} ^{\,\,\,a}$ in terms of the
fundamental representation $M^{n} _{\,\,\,m}$ we define a right coinvariant
Maurer-Cartan 1-form $\omega $ on $SO_{q}\left( 6\right) $ as 
\begin{equation}
\omega ^{n} _{\,\,\,k}=dM^{n} _{\,\,\,m}S\left( M^{m} _{\,\,\,\,\,k}\right)
\end{equation}
and 
\begin{equation}
\tilde{\theta }^{n} _{\,\,\,k}=S\left( M^{n} _{\,\,\,m}\right) dM^{m}
_{\,\,\,\,\,k} .
\end{equation}
The left coinvariant Maurer-Cartan 1-forms $\tilde{ \theta }$ and the right
coinvariant Maurer-Cartan 1-form $\omega $ are related by

\begin{eqnarray}
\tilde{\theta }^{i} _{\,\,\,j}&=&S\left( M^{i} _{\,\,\,m}\right) \omega ^{m}
_{\,\,\,\,\,\,k}M^{k} _{\,\,\,j}  \notag \\
\omega ^{i} _{\,\,\,j}&=&M^{i} _{\,\,\,m}\theta ^{m} _{\,\,\,\,\,\,k}S\left(
M^{k} _{\,\,\,j}\right) .
\end{eqnarray}

Now we calculate the right coaction on $\tilde{\theta }^{i} _{\,\,\,j}$ by
expressing them in terms of $\omega ^{i} _{\,\,\,j}$, using the homomorphism
property of the right coaction, the right coinvariance of $\omega ^{i}
_{\,\,\,j}$ and translating the latter back into the $\tilde{\theta }^{i}
_{\,\,\,j}$. The result is 
\begin{eqnarray}
\Delta _{R}\left( \tilde{\theta }^{i} _{\,\,\,j}\right) &=&\tilde{\theta }%
^{l} _{\,\,\,n}\otimes S\left( M^{i} _{\,\,\,l}\right) M^{n} _{\,\,\,j} 
\notag \\
&=&\tilde{\theta }^{l} _{\,\,\,n}\otimes M_{l\,\,\,\,\,\,\,\,\,\,j}
^{\,\,\,ni}
\end{eqnarray}
with 
\begin{equation}
M_{l\,\,\,\,\,\,\,\,\,\,j} ^{\,\,\,ni}=S\left(M^{i} _{\,\,\,l}\right)M^{n}
_{\,\,\,j} .
\end{equation}
If we replace the index pairs $^{i} _{\,\,\,j}$ with the upper indices (row
indices) $a=1,...,36$ and the index pairs $_{l} ^{\,\,\,n}$ with lower
indices (column indices) $b$, Equ. (92) gives 
\begin{equation}
\Delta _{R}\left( \tilde{\theta }^{a}\right)\tilde{\theta }^{b}\otimes M_{b}
^{\,\,\,a}.
\end{equation}
The co-structures on $M_{b} ^{\,\,\,a}$ are given by 
\begin{eqnarray}
\Delta \left( M_{b} ^{\,\,\,a}\right) &=& M_{b} ^{\,\,\,c}\otimes M_{c}
^{\,\,\,a} ,  \notag \\
\epsilon \left( M_{b} ^{\,\,\,a}\right) &=& \delta ^{a} _{b} ,  \notag \\
S\left( M_{b} ^{\,\,\,c}\right) M_{c} ^{\,\,\,a} &=& M_{b} ^{\,\,\,c}M_{c}
^{\,\,\,a}=\delta ^{a} _{b} .
\end{eqnarray}
In the quantum case we have $\tilde{\theta }^{i} _{\,\,\,j}M^{n}
_{\,\,\,m}\neq M^{n} _{\,\,\,m}\tilde{\theta }^{i} _{\,\,\,j}$ in general,
the bimodule structure of $\Gamma $ being non-trivial for $q\neq 1$. There
exist linear functionals $f^{i\,\,\,\,\,\,\,\,p} _{\,\,\,jq} : Fun\left(
SO_{q}\left( 6\right) \right) \rightarrow \mathbb{C}$ for these left
invariant basis such that

\begin{eqnarray}
\tilde{\theta }^{i} _{\,\,\,j}M^{n} _{\,\,\,m} &=& \left(
f^{i\,\,\,\,\,\,\,\,q} _{\,\,\,jp}\star M^{n} _{m}\right) \tilde{\theta }%
^{p} _{\,\,\,q} = \left( id \otimes f^{i\,\,\,\,\,\,\,\,q}
_{\,\,\,jp}\right) \Delta \left( M^{n} _{\,\,\,m}\right) \tilde{\theta }^{p}
_{q}  \notag \\
&=& M^{n} _{\,\,\,\,l}\,\, f^{i\,\,\,\,\,\,\,\,q} _{\,\,\,jp}\left( M^{l}
_{\,\,\,m} \right) \tilde{\theta }^{p} _{\,\,\,q} ,
\end{eqnarray}
\begin{equation}
M^{n} _{\,\,\,m}\tilde{\theta }^{i} _{\,\,\,j}=\tilde{\theta }^{p} _{\,\,\,q}%
\left[ \left( f^{i\,\,\,\,\,\,\,\,q} _{\,\,\,jp}\circ S\right) \star M^{n}
_{\,\,\,m}\right] .
\end{equation}

The functionals $f^{i\,\,\,\,\,\,\,\,q} _{\,\,\,jp}$ are uniquely determined
by Equ. (96) and satisfy 
\begin{eqnarray}
f^{i\,\,\,\,\,\,\,\,q} _{\,\,\,jp}\left( M^{n} _{\,\,\,m}M^{l}
_{\,\,\,k}\right) &=& f^{i\,\,\,\,\,\,\,\,s} _{\,\,\,jr}\left( M^{n}
_{\,\,\,m}\right) f^{r\,\,\,\,\,\,\,\,q} _{\,\,\,sp}\left( M^{l}
_{\,\,\,k}\right) \\
\left( f^{i\,\,\,\,\,\,\,\,q} _{\,\,\,jp}\star M^{n} _{\,\,\,m}\right)
M^{\,\,\,\,sp} _{r\,\,\,\,\,\,\,\,\,\,q} &=& M^{\,\,\,\,pi}
_{q\,\,\,\,\,\,\,\,\,\,j}\left( M^{n} _{\,\,\,m}\star f^{q\,\,\,\,\,\,\,\,s}
_{\,\,\,pr} \right) .
\end{eqnarray}
This result places significant constraints on the possible bicovariant
calculi which are consistent with the assumption that the differentials of
the generators should generate the bimodules of forms as a left $\mathcal{A}$%
-module. In fact we are passing directly to a class of quotients of the
bimodule $\Gamma $ which we then constraint by the requirement that the
bicovariance is not destroyed. \newline
We can define a new basis

\begin{equation}
\tilde{\theta }^{nm}=\tilde{\theta }^{m} _{\,\,\,j}C^{jn}, \quad \quad 
\tilde{\theta }^{m} _{\,\,\,n}=\tilde{\theta }^{mj}C_{jn} .
\end{equation}
The functionals $f^{ml} _{\,\,\,\,\,\,ij}$ corresponding to the basis $%
\tilde{\theta }^{mn}$ are given by 
\begin{equation}
f^{ml} _{\,\,\,\,\,\,\,\,ij}=C^{pl} f^{m\,\,\,\,\,\,\,\,q}
_{\,\,\,\,\,\,pi}C_{jq} .
\end{equation}
The relation between the left and the right multiplication for this basis is 
\begin{equation}
\tilde{\theta }^{mn}M^{i} _{\,\,\,j}=\left( f^{mn}
_{\,\,\,\,\,\,\,\,pq}\star M^{i} _{\,\,\,j}\right) \tilde{\theta }^{pq} .
\end{equation}
The exterior derivative $d$ is defined as 
\begin{eqnarray}
dM^{n} _{\,\,\,m}&=&\frac{1}{\mathcal{N}}\left[ \tilde{X}, M^{n} _{\,\,\,m}%
\right] = \left( \tilde{\chi }^{\,\,\,j} _{i} \star M^{n} _{\,\,\,m}\right) 
\tilde{\theta }^{i} _{\,\,\,j}  \notag \\
&=& \left( id \otimes \tilde{\chi }^{\,\,\,j} _{i}\right) \Delta \left(
M^{n} _{\,\,\,m}\right) \tilde{\theta }^{i} _{\,\,\,j}= \tilde{\chi }%
^{\,\,\,j} _{i}\left( M^{k} _{\,\,\,m}\right) M^{n} _{\,\,\,k}\tilde{\theta }%
^{i} _{\,\,\,j} ,
\end{eqnarray}
where $\tilde{X}=\tilde{\theta }^{i}_{\,\,\,i}$ is the singlet
representation of $\tilde{\theta }^{i} _{\,\,\,j}$, $\mathcal{N}\in \mathbb{C%
}$ is the normalization constant and $\tilde{\chi }^{\,\,\,j} _{i}$ are
left-invariant vector fields. \newline
Now, let us find the relation between the left Cartan-Maurer forms $\tilde{%
\theta }^{i} _{\,\,\,j}$ and the forms obtained by taking the tensor product 
$\theta ^{ab}=\theta ^{a} \bar{\theta }^{b} $. \newline
From Equ. (90) and Equ. (66) we get 
\begin{equation}
\tilde{\theta }^{i} _{\,\,\,j}=\chi _{ab}\left( M^{i} _{\,\,\,j}\right)
\theta ^{ab}.
\end{equation}
Equ. (64) gives 
\begin{equation}
\chi _{ab}\left( M^{i} _{\,\,\,j}\right) \theta ^{ab}M^{n} _{\,\,\,m}=f^{ab}
_{Ad\,\,\,\,\,\,cd}\left( M^{k} _{\,\,\,m}\right) M^{n} _{\,\,\,k}\chi
_{ab}\left( M^{i} _{\,\,\,j}\right) \theta ^{cd}.
\end{equation}
Comparing with Equ. (96) we find

\begin{equation}
f^{cd} _{Ad\,\,\,\,\,\,ab} \left( M^{k} _{\,\,\,m}\right) \chi _{cd} \left(
M^{i} _{\,\,\,j}\right) = f^{i\,\,\,\,\,\,\,\,q} _{\,\,\,\,jp}\left( M^{k}
_{\,\,\,m}\right) \chi _{ab} \left( M^{p} _{\,\,\,q} \right) .
\end{equation}
This is the first identity. The second identity gives a relation between
left-invariant vector fields. In fact, from Equ. (66) and Equ. (103) we get

\begin{eqnarray}
dM^{n} _{\,\,\,m} &=& M^{n} _{\,\,\,k}\chi _{ab}\left( M^{k}
_{\,\,\,m}\right) \theta ^{ab}  \notag \\
&=& \tilde{\chi }^{\,\,\,j} _{i}\left( M^{k} _{\,\,\,m}\right) M^{n}
_{\,\,\,k} \tilde{\theta }^{i} _{\,\,\,j} = \tilde{\chi }^{\,\,\,j}
_{i}\left( M^{k} _{\,\,\,m}\right) M^{n} _{\,\,\,k}\chi _{ab}\left( M^{i}
_{\,\,\,j}\right) \theta ^{ab},
\end{eqnarray}

which gives

\begin{equation}
\chi _{ab}\left( M^{k} _{\,\,\,m}\right) = \tilde{\chi }^{\,\,\,j}
_{i}\left( M^{k} _{\,\,\,m}\right) \chi _{ab}\left( M^{i} _{\,\,\,j}\right) .
\end{equation}

Taking $\tilde{\chi }^{\,\,\,j} _{i}\left( M^{k} _{\,\,\,m}\right) =\delta
^{j} _{m}\delta ^{k} _{i} $ we find an identity.

\section{Quantum Gauge Theories}

We formulate the $q$-deformed gauge theories using the concept of the
quantum fiber bundles \cite{bernard, mesref3}. We consider a quantum vector $%
E\left( X_{B},V,\mathcal{A}\right) $ with a noncommutative algebra base $%
X_{B}$ (the quantum spacetime), a comodule algebra $V$ considered as a fiber
of $E$ and a structure quantum group $\mathcal{A}$ playing the role of the
quantum symmetry group. The matter fields (sections) $\psi $ are maps: $%
V\rightarrow X_{B}$. The quantum gauge transformations $T$ are defined in
terms of the right coaction and act on the sections as

\begin{eqnarray}
\psi ^{\prime i} &=& \left( \psi \star T\right) \left( \theta ^{i}\right)
=\left( \psi \otimes T\right) \Delta _{R}\left( \theta ^{i}\right)  \notag \\
&=&\psi \left( \theta ^{j}\right) \otimes T\left( S\left( M^{i}
_{\,\,\,j}\right) \right) = \psi \left( \theta ^{j}\right) \otimes
T^{-1}\left( M^{i} _{\,\,\,j}\right),
\end{eqnarray}

where $\theta ^{i}\in V$ and $T: \mathcal{A}\rightarrow X_{B}$ is a
convolution invertible map such that $T\left( 1_{\mathcal{A}}\right)
=1_{X_{B}}$.

For any two quantum gauge transformations $T$ and $T^{\prime }$, the
convolution product is defined as 
\begin{equation}
\left( T\star T^{\prime }\right) \left( M^{i} _{\,\,\,j}\right) = \left(
T\otimes T^{\prime }\right) \Delta \left( M^{i} _{\,\,\,j}\right) .
\end{equation}

The quantum inverse gauge transformation is defined as 
\begin{equation}
T^{-1 i} _{\,\,\,\,\,\,\,\,\,\,\,\,\,j}=T\left( S\left( M^{i}
_{\,\,\,\,j}\right) \right) .
\end{equation}
Using the right covariance property of the quantum $\mathcal{A}$-bimodule $V$%
, we get 
\begin{eqnarray}
\psi ^{\prime \prime i} &=& \left( \psi ^{\prime }\star T\right) \left(
\theta ^{i}\right) = \left( \left( \psi \star T \right) \star \right) \left(
\theta ^{i}\right)  \notag \\
&=& \left( \psi \otimes T \otimes T \right) \left( id\otimes \Delta \right)
\Delta _{R}\left( \theta ^{i}\right)  \notag \\
&=& \left( \psi \star \left( T\star T\right) \right) \left( \theta
^{i}\right)
\end{eqnarray}
which simply reflect the closure of the finite quantum gauge
transformations. \newline
Let us now define a covariant exterior derivative as a linear map on the set
of sections $\Gamma \left( E\right) $, $\nabla : \Gamma \left( E \right)
\rightarrow \Gamma ^{1}\left( E \right) $, where $\Gamma ^{1}\left( E\right) 
$ is the set of one-form sections. We require that these sections transform
with the same rule as the corresponding matter fields, i.e. : 
\begin{eqnarray}
\nabla ^{\prime n} _{\,\,\,\,\,\,m} &=& \left( \nabla \otimes T\right)
Ad_{R}\left( M^{n} _{\,\,\,\,\,m}\right)  \notag \\
&=& \nabla ^{k} _{\,\,\,\,l}\otimes T^{-1 n}
_{\,\,\,\,\,\,\,\,\,\,\,\,\,\,k}T^{l} _{\,\,\,\,m},
\end{eqnarray}
which gives 
\begin{eqnarray}
\nabla ^{\prime n} _{\,\,\,m}\psi ^{\prime m} &=& \left( \nabla ^{k}
_{\,\,\,l} \otimes T^{-1 n} _{\,\,\,\,\,\,\,k}T^{l}
_{\,\,\,\,\,\,\,m}\right) \left( \psi ^{j} \otimes T^{-1 m}
_{\,\,\,\,\,\,\,j}\right)  \notag \\
&=& \nabla ^{k} _{\,\,\,j}\psi ^{j}\otimes T^{-1 n}
_{\,\,\,\,\,\,\,\,\,\,\,\,\,\,k} .
\end{eqnarray}

The quantum exterior derivative $d$ on the base space $X_{B}$ is defined in
terms of the covariant derivative 
\begin{equation}
\nabla ^{n} _{\,\,\,m}=d\delta ^{n} _{m}+A^{n} _{\,\,\,m}
\end{equation}
where $A^{n} _{\,\,\,m}$ are the quantum Lie-algebra valued matrices of one
forms on $X_{B}$, i.e. $A^{n} _{\,\,\,m}=A^{ab}\chi _{ab}\left( M^{n}
_{\,\,\,m}\right) $. \newline

The consistency of Equ. (113) requires that $A^{n} _{\,\,\,m}$ transforms as 
\begin{equation}
A^{\prime n} _{\,\,\,\,\,\,\,\,\,\,m}=A^{k} _{\,\,\,l}\otimes T^{-1 n}
_{\,\,\,\,\,\,\,\,\,\,\,\,k}T^{l} _{\,\,\,m}+\delta ^{k} _{l}\otimes T^{-1
n} _{\,\,\,\,\,\,\,\,\,\,\,\,k}dT^{l} _{\,\,\,m} ,
\end{equation}
and two successive transformations act as 
\begin{eqnarray}
A^{\prime n} _{\,\,\,\,\,\,\,\,m} &=& A^{\prime k}
_{\,\,\,\,\,\,\,\,l}\otimes T^{-1 n} _{\,\,\,\,\,\,\,\,\,\,\,\,k}T^{l}
_{\,\,\,m}+\delta ^{k} _{l}\otimes T^{-1 n}
_{\,\,\,\,\,\,\,\,\,\,\,\,k}dT^{l} _{\,\,\,m}  \notag \\
&=& A^{p} _{\,\,\,q}\otimes T^{-1 k} _{\,\,\,\,\,\,\,\,\,\,\,\,p}T^{q}
_{\,\,\,l}\otimes T^{-1 n} _{\,\,\,\,\,\,\,\,\,\,\,\,k}T^{l} _{\,\,\,m} 
\notag \\
& &\,\, +1\otimes T^{-1 k} _{\,\,\,\,\,\,\,\,\,\,\,\,p}dT^{p}
_{\,\,\,l}\otimes T^{-1 n} _{\,\,\,\,\,\,\,\,\,\,\,\,k}T^{l}
_{\,\,\,m}+1\otimes T^{-1 n} _{\,\,\,\,\,\,\,\,\,\,\,\,p}dT^{p} _{m}  \notag
\\
&=& A^{p} _{\,\,\,q}\otimes \left( T^{n} _{\,\,\,k}\otimes T^{k}
_{\,\,\,p}\right) ^{-1}\left( T^{q} _{\,\,\,l}\otimes T^{l} _{\,\,\,m}\right)
\notag \\
& &\,\, +1\otimes \left( T^{-1 k} _{\,\,\,\,\,\,\,\,\,\,\,\,p}\otimes T^{-1
n} _{\,\,\,\,\,\,\,\,\,\,\,\,k}\right) \left( dT^{p} _{\,\,\,l} \otimes
T^{l} _{\,\,\,m}\right)+1\otimes T^{-1 n} _{\,\,\,\,\,\,\,\,\,\,\,\,p}dT^{p}
_{\,\,\,m}  \notag \\
&=& A^{p} _{\,\,\,q}\otimes \left( T^{n} _{\,\,\,k}\otimes T^{k}
_{\,\,\,p}\right) ^{-1}\left( T^{q} _{\,\,\,l}\otimes T^{l} _{\,\,\,m}\right)
\notag \\
& &\,\, +1\otimes \left( T^{n} _{\,\,\,k}\otimes T^{k} _{\,\,\,p}\right)
^{-1}d\left( T^{p} _{\,\,\,l}\otimes T^{l} _{\,\,\,m}\right) .
\end{eqnarray}
This equation shows that if $T^{n} _{\,\,\,m}$ is a gauge transformation on
the connection $A^{p} _{\,\,\,q}$, then $T^{\prime n} _{\,\,\,\,m}=\left( T
\star T\right) \left( M^{n} _{\,\,\,m}\right) =\left( T\otimes T\right)
\Delta \left( M^{n} _{\,\,\,m}\right) $. \newline

As in the undeformed case, the quantum two-form curvature associated to the
connection is given by 
\begin{equation}
F^{n} _{\,\,\,m}=\nabla ^{n} _{\,\,\,k}\wedge \nabla ^{k} _{\,\,\,m}=dA^{n}
_{\,\,\,m}+A^{n} _{\,\,\,k}\wedge A^{k} _{\,\,\,m} ,
\end{equation}
and transforms as 
\begin{eqnarray}
F^{\prime n} _{\,\,\,m} &=& \nabla ^{\prime n} _{\,\,\,\,\,i}\wedge \nabla
^{\prime i} _{\,\,\,\,\,m}=\left( \nabla ^{j} _{\,\,\,l}\otimes T^{-1 n}
_{\,\,\,\,\,\,\,\,\,\,\,j}T^{l} _{\,\,\,i}\right) \left( \nabla ^{k}
_{\,\,\,p}\otimes T^{-1 i} _{\,\,\,\,\,\,\,\,\,\,\,k}T^{p} _{\,\,\,m}\right)
\notag \\
&=& F^{j} _{\,\,\,k}\otimes T^{-1 n} _{\,\,\,\,\,\,\,\,\,\,\,\,\,j}T^{k}
_{\,\,\,m} .
\end{eqnarray}

We can also express this equation in terms of the right adjoint coaction as 
\begin{equation}
F^{\prime n} _{\,\,\,\,\,m}=F^{\prime }\left( M^{n} _{\,\,\,m}\right)
Ad_{R}\left( M^{n} _{\,\,\,m}\right) = \left( F\wedge T\right) \left( M^{n}
_{\,\,\,m}\right) .
\end{equation}

The closure of the gauge transformations for the curvature can be obtained
as 
\begin{eqnarray}
F^{\prime \prime }\left( M^{n} _{\,\,\,m}\right) &=& \left( F^{\prime
}\wedge T\right) \left( M^{n} _{\,\,\,m} \right)=\left( \left( F\wedge
T\right) \wedge T\right) \left( M^{n} _{\,\,\,m}\right)  \notag \\
&=& F^{i} _{\,\,\,j}\otimes \left( T^{n} _{\,\,\,k}\otimes T^{k}
_{\,\,\,i}\right) ^{-1}\left( T^{j} _{\,\,\,l}\otimes T^{l}
_{\,\,\,m}\right) .
\end{eqnarray}

We can decompose the curvature in terms of left-invariant vector basis as 
\begin{eqnarray}
F^{n} _{\,\,\,m} &=& F^{ab}\chi _{ab}\left(M^{n} _{\,\,\,m}\right)  \notag \\
&=& dA^{ab}\chi _{ab}\left(M^{n} _{\,\,\,m}\right) + A^{ab}\wedge A^{cd}\chi
_{ab}\left(M^{n} _{\,\,\,p}\right) \chi _{cd}\left(M^{p} _{\,\,\,m}\right) .
\end{eqnarray}
We define the infinitesimal variations around unity of the gauge
transformations as 
\begin{eqnarray}
\delta _{\alpha }T^{n} _{\,\,\,m} &=& \left( T \star \alpha \right) \left(
M^{n} _{\,\,\,m}\right) = \left( T \star \alpha ^{ab} \chi _{ab}\right)
\Delta \left( M^{n} _{\,\,\,m} \right)  \notag \\
&=& T^{n} _{\,\,\,k}\otimes \alpha ^{ab} \chi _{ab}\left( M^{k} _{\,\,\,m}
\right) ,
\end{eqnarray}
where $\alpha $ are infinitesimal quantum gauge parameters of the
transformation $T$. The infinitesimal variation corresponding to $\alpha $
is given by

\begin{equation}
\alpha ^{\prime }\left( M^{n} _{\,\,\,m}\right) = \left( \alpha \otimes
T\right) Ad_{R}\left( M^{n} _{\,\,\,m}\right) .
\end{equation}
The infinitesimal variation of the connection is 
\begin{equation}
\delta _{\alpha }A^{ab}=-1\otimes d\alpha ^{ab}+A^{cd}\otimes \alpha ^{ef}
C^{\,\,\,\,\,\,\,\,\,\,\,\,ab} _{cdef} ,
\end{equation}
where $C^{\,\,\,\,\,\,\,\,\,\,\,\,ab} _{cdef}$ are the quantum structure
constants defined in Equ. (80).

The curvature transforms with the right adjoint coaction. Its infinitesimal
gauge transformation reads 
\begin{equation}
\delta _{\alpha }F^{n} _{\,\,\,m}=\left( F\otimes \alpha \right)
Ad_{R}\left( M^{n} _{\,\,\,m}\right) .
\end{equation}
In terms of components, we find 
\begin{equation}
\delta _{\alpha }F^{ab}=F^{cd}\otimes \alpha
^{ef}C^{\,\,\,\,\,\,\,\,\,\,\,\,ab} _{cdef} .
\end{equation}

To illustrate the construction of BRST and anti-BRST transformations, let us
consider the simple example of BF-Yang-Mills theories. These theories supply
a complete nonperturbative Nicolai map for Yang-Mills theory on any Riemann
surface which reduces the partition function to an integral over the moduli
space of flat connections, with measure given by the Ray-Singer torsion. 
\newline

The quantum Lagrangian describing these models (hereafter, the spacetime
indices are omitted for simplicity) can be written as 
\begin{equation}
L_{BFYM}=< iB^{ab}F^{cd}+g^{2}B^{ab}B^{cd} >g_{abcd}
\end{equation}
where $B$ is a quantum Lie-algebra valued 2-form, $g^{2}$ is the coupling
constant and $g_{abcd}$ is the quantum Killing metric defined in Equ. (83).
The quantum Lie-algebra valued curvature $F: \mathcal{A}\rightarrow \Gamma
^{2}\left( X_{B}\right) $ is given by Equ. (118). \newline
The quantum BRST transformations for these models are obtained, as usual, by
replacing the quantum infinitesimal parameters by the ghosts:

\begin{eqnarray}
\delta A &=&-dc^{ab}\chi _{ab}-A^{ab}\cdot c^{cd}\left( \chi _{ab}\otimes
\chi _{cd}\right) Ad_{R},  \notag \\
\delta F &=&-F^{ab}\cdot c^{cd}\left( \chi _{ab}\otimes \chi _{cd}\right)
Ad_{R},  \notag \\
\delta B &=&-B^{ab}\cdot c^{cd}\left( \chi _{ab}\otimes \chi _{cd}\right)
Ad_{R},  \notag \\
\delta c &=&-\frac{1}{2}c^{ab}\cdot c^{cd}\left( \chi _{ab}\otimes \chi
_{cd}\right) Ad_{R},  \notag \\
\delta \overline{c} &=&b,\quad \delta b=0
\end{eqnarray}

\bigskip

and the corresponding quantum anti-BRST transformations as

\bigskip

\begin{eqnarray}
\overline{\delta }A &=&-d\overline{c}^{ab}\chi _{ab}-A^{ab}\cdot
c^{cd}\left( \chi _{ab}\otimes \chi _{cd}\right) Ad_{R},  \notag \\
\overline{\delta }F &=&-F^{ab}\cdot \overline{c}^{cd}\left( \chi
_{ab}\otimes \chi _{cd}\right) Ad_{R},  \notag \\
\overline{\delta }B &=&-B^{ab}\cdot \overline{c}^{cd}\left( \chi
_{ab}\otimes \chi _{cd}\right) Ad_{R},  \notag \\
\overline{\delta }c &=&-b-\frac{1}{2}c^{ab}\cdot c^{cd}\left( \chi
_{ab}\otimes \chi _{cd}\right) Ad_{R},  \notag \\
\overline{\delta }\overline{c} &=&-\frac{1}{2}\overline{c}^{ab}\cdot 
\overline{c}^{cd}\left( \chi _{ab}\otimes \chi _{cd}\right) Ad_{R},  \notag
\\
\overline{\delta }b &=&-b^{ab}\cdot \overline{c}^{cd}\left( \chi
_{ab}\otimes \chi _{cd}\right) Ad_{R}.
\end{eqnarray}

\bigskip

A straightforward but tedious calculation using essentially the quantum
Jacobi identity shows that the quantum BFYM action is separately invariant
under these quantum BRST and anti-BRST transformations. Using the quantum
BRST transformations we also find the quantum analogue of the Taylor-Slavnov
identity.

\bigskip

To define the quantum analogue of the Batalin-Vilkovisky operator \cite%
{batalin} we have to introduce the quantum left and right functional \
partial derivatives. Given a quantum field $\varphi =\varphi ^{ab}\chi _{ab} 
$, $\varphi ^{\ast }=\varphi ^{\ast ab}\chi _{ab}$ and a quantum functional $%
F$ we define

\bigskip

\begin{equation}
\frac{d}{dt}{\LARGE \mid }_{t=0}\,\,F\left( \varphi ^{ab}+t\varrho
^{ab}\right) =\int_{X_{B}}\left\langle \varrho ^{ab},\frac{\overrightarrow{%
\partial }F}{\partial \varphi _{cd}}\right\rangle
\,g_{abcd}=\int_{X_{B}}\left\langle \frac{F\overleftarrow{\partial }}{%
\partial \varphi _{ab}},\varrho ^{cd}\right\rangle \,g_{abcd}
\end{equation}

and the quantum Batalin Vilkovisky antibracket of two functionals $F$, $G$:

\begin{equation}
\left( F,G\right) = \int_{X_{B}}\left\langle \frac{F\overleftarrow{\partial }%
}{\partial \varphi ^{i} _{ab}}, \frac{\overrightarrow{\partial }G}{\partial
\varphi ^{+} _{i\, cd}}\right\rangle g_{abcd} \, - \left( -1\right) ^{%
\mathrm{deg} \varphi ^{i\, ab}}\left\langle \frac{F\overleftarrow{\partial }%
}{\partial \varphi ^{+\,i} _{ab}}, \frac{\overrightarrow{\partial }G}{%
\partial \varphi _{i\, cd}}\right\rangle g_{abcd} .
\end{equation}

\bigskip

We can construct a new action called the quantum BRST action defined by

\begin{equation}
S=\int_{X_{B}}\,\,d^{4} _{q}x \left( L_{BFYM}+\left( -\right) ^{\epsilon
_{i}}\varphi ^{\ast \,ab} _{i}\,\delta \,\varphi ^{i \,cd} g_{abcd}\right) ,
\end{equation}

where we have used the quantum BRST transformations $\delta $ defined in
Equ. (129). $\epsilon _{i}$ is the Grassmann parity of $\varphi ^{i}$ and $%
g_{abcd}$ is the quantum Killing metric. We have denoted all the antifields
by $\varphi ^{\ast \,i \, ab}$ and the fields and the ghosts by $\varphi ^{i
\, ab}$. \newline
By construction the quantum BRST action is such:

\begin{eqnarray}
\frac{\overrightarrow{\partial }S}{\delta \varphi ^{\ast \, i \, ab}} &=&
-\delta \varphi ^{i \, cd}\, g_{abcd}= - \left( S, \varphi ^{i\, cd}\right)
g_{abcd}  \notag \\
\frac{\overrightarrow{\partial }S}{\delta \varphi ^{i \, ab}} &=& -\delta
\varphi ^{\ast \, i \, cd}\, g_{cdab}=- \left( S, \varphi ^{\ast \, i\,
cd}\right) g_{cdab} .
\end{eqnarray}

These quantum functionals derivatives satisfy the quantum Jacobi identities.
Furthermore, the quantum action $S$ satisfies the quantum master equation

\begin{equation}
\left( S, S\right) =0 ,
\end{equation}

which is a direct consequence of $\delta ^{2}=0$. \newline
Now, we define the quantum Batalin Vilkovisky operator as

\begin{equation}
\Delta = \left( - \right) ^{\epsilon _{i}} \frac{\overrightarrow{\partial }}{%
\partial \varphi ^{i} _{ab}}\frac{\overrightarrow{\partial }}{\partial
\varphi ^{\ast } _{i \, cd}} g_{abcd} .
\end{equation}

The quantum Batalin Vilkovisky operator coincides with the usual one in the
limit $q=1$.

\section{A Map between $q$-Gauge Fields and Ordinary Gauge Fields}

The starting point for this investigation is the wish to define the
analogue, in the quantum group picture \cite{drinfeld}, of the Seiberg
Witten map \cite{seiberg} argued using the ideas of noncommutative geometry 
\cite{connes}. In the framework of string theory Seiberg and Witten have
noticed that the noncommutativity depends on the choice of the
regularization procedure: it appears in point-spliting regularization
whereas it is not present in the Pauli Villars regularization. This
observation led them to argue that there exists a map connecting the
noncommutative gauge fields and gauge transformation parameter to the
ordinary gauge fields and gauge parameter. This map can be interpreted as an
expansion of the noncommutative gauge field in $\theta $. \ Along similar
lines, we have introduced in Refs \cite{mesref5, mesref6, mesref7} a new map
between the $q$-deformed and undeformed gauge theories. This map can be seen
as an infinitesimal shift in the parameter $q$, and thus as an expansion of
the deformed gauge fields in $q$.

To begin we consider the undeformed action 
\begin{equation}
S=-\frac{1}{4}\int d^{4}x\,\,F_{\mu \nu }F^{\mu \nu },
\end{equation}
where

\begin{equation}
F_{\mu \nu }=\partial _{\mu }A_{\nu }-\partial _{\nu }A_{\mu }.
\end{equation}

$S$ is invariant with respect to infinitesimal gauge transformation:

\begin{equation}
\delta _{\lambda }A_{\mu }=\partial _{\mu }\lambda .
\end{equation}

Now let us study the quantum gauge theory on the quantum plane $\hat{x}\hat{y%
}=q\hat{y}\hat{x}$. In general, the product of functions on a deformed space
is defined via the Gerstenhaber star product \cite{gerstenhaber}: Let $%
\mathcal{A}$ be an associative algebra and let $D_{i},E^{i}:\mathcal{A}%
\rightarrow \mathcal{A}$ be a pairwise derivations. \newline
Then the star product of $a$ and $b$ is given by

\begin{equation}
a\star b=\mu \circ e^{\zeta \sum_{i}D_{i}\otimes E^{i}}a\otimes b,
\end{equation}

where $\zeta $ is a parameter and $\mu $ the undeformed product given by

\begin{equation}
\mu \left( f\otimes g\right) =fg.
\end{equation}

On the Manin plane $\hat{x}\hat{y}=q\hat{y}\hat{x}$, we can write this star
product as:

\begin{equation}
f\star g=\mu \circ e^{\frac{i\eta }{2}\left( x\frac{\partial }{\partial x}%
\otimes y\frac{\partial }{\partial y}-y\frac{\partial }{\partial y}\otimes x%
\frac{\partial }{\partial x}\right) }f\otimes g
\end{equation}

A straightforward computation gives then the following commutation relations

\begin{equation}
x\star y=e^{\frac{i\eta }{2}}xy,\qquad \qquad y\star x=e^{\frac{-i\eta }{2}
}yx.
\end{equation}

Whence

\begin{equation}
x\star y=e^{i\eta }y\star x,\qquad q=e^{i\eta }.
\end{equation}

Thus we recover the commutation relations for the Manin plane: \newline
$\hat{x}\hat{y}=q\hat{y}\hat{x}$. \newline
We can also write the product of functions as

\begin{equation}
f\star g=fe^{\frac{i}{2}\overleftarrow{\partial }_{k}\theta ^{kl}\left(
x,y\right) \overrightarrow{\partial }_{l}}g
\end{equation}

where the antisymmetric matrix $\theta ^{kl}\left( x,y\right) =\eta
xy\epsilon ^{kl}$ \newline
with $\epsilon ^{12}=-\epsilon ^{21}=1$. \newline

Expanding to first nontrivial order in $\eta $, we find

\begin{equation}
f\star g = fg+\frac{i}{2}\eta xy \left( \frac{\partial f}{\partial x}\frac{%
\partial g}{\partial y}-\frac{\partial f}{\partial y}\frac{\partial g}{%
\partial x}\right) .
\end{equation}

The $q$-deformed infinitesimal gauge transformations are defined by

\begin{eqnarray}
\widehat{\delta }_{\widehat{\lambda }}\widehat{A}_{\mu } &=&\partial _{\mu }%
\widehat{\lambda }+i\left[ \widehat{\alpha },\widehat{A}_{\mu }\right]
_{\star }=\partial _{\mu }\widehat{\lambda }+i\widehat{\lambda }\star 
\widehat{A}_{\mu }-i\widehat{A}_{\mu }\star \widehat{\lambda },  \notag \\
\widehat{\delta }_{\widehat{\lambda }}\widehat{F}_{\mu \nu } &=&i\widehat{%
\lambda }\star \widehat{F}_{\mu \nu }-i\widehat{F}_{\mu \nu }\star \widehat{%
\lambda }.
\end{eqnarray}

To first order in $\theta \left( x,y\right) $, the above formulas for the
gauge transformations read

\begin{eqnarray}
\widehat{\delta }_{\widehat{\lambda }}\widehat{A}_{\mu } &=&\partial _{\mu }%
\widehat{\lambda }-\frac{1}{2}\theta ^{\rho \sigma }\left( x,y\right) \left(
\partial _{\rho }\lambda \partial _{\sigma }A_{\mu }-\partial _{\rho }A_{\mu
}\partial _{\sigma }\lambda \right)  \notag \\
\widehat{\delta }_{\widehat{\lambda }}\widehat{F}_{\mu \nu } &=&-\frac{1}{2}%
\theta ^{\rho \sigma }\left( x,y\right) \left( \partial _{\rho }\lambda
\partial _{\sigma }F_{\mu \nu }-\partial _{\rho }F_{\mu \nu }\partial
_{\sigma }\lambda \right) .
\end{eqnarray}

To ensure that an ordinary gauge transformation of $A$ by $\lambda $ is
equivalent to $q$-deformed gauge transformation of $\widehat{A}$ by $%
\widehat{\lambda }$ we consider the following relation \cite{seiberg}

\begin{equation}
\widehat{A}\left( A\right) +\widehat{\delta }_{\widehat{\lambda }}\widehat{A}%
\left( A\right) =\widehat{A}\left( A+\delta _{\lambda }A\right)
\end{equation}

We first work the first order in $\theta $

\begin{eqnarray}
\widehat{A} &=&A+A^{\prime }\left( A\right)  \notag \\
\widehat{\lambda }\left( \lambda ,A\right) &=&\lambda +\lambda ^{\prime
}\left( \lambda ,A\right) .
\end{eqnarray}

Expanding in powers of $\theta $ we find

\begin{equation}
A_{\mu }^{\prime }\left( A+\delta _{\lambda }A\right) -A_{\mu }^{\prime
}\left( A\right) -\partial _{\mu }\lambda ^{\prime }=\theta ^{kl}\left(
x,y\right) \partial _{k}A_{\mu }\partial _{l}\lambda
\end{equation}

The solutions are given by

\begin{eqnarray}
\widehat{A}_{\mu } &=&A_{\mu }-\frac{1}{2}\theta ^{\rho \sigma }\left(
x,y\right) \left( A_{\rho }F_{\sigma \mu }+A_{\rho }\partial _{\sigma
}A_{\mu }\right) , \\
\widehat{\lambda } &=&\lambda +\frac{1}{2}\theta ^{\rho \sigma }\left(
x,y\right) A_{\sigma }\partial _{\rho }\lambda .
\end{eqnarray}

The $q$-deformed curvature $\widehat{F}_{\mu \nu }$ is given by

\begin{eqnarray}
\widehat{F}_{\mu \nu } &=&\partial _{\mu }\widehat{A}_{\nu }-\partial _{\nu }%
\widehat{A}_{\mu }-i\left[ \widehat{A}_{\mu },\widehat{A}_{\nu }\right]
_{\star }  \notag \\
&=&\partial _{\mu }\widehat{A}_{\nu }-\partial _{\nu }\widehat{A}_{\mu }-i%
\widehat{A}_{\mu }\star \widehat{A}_{\nu }+i\widehat{A}_{\nu }\star \widehat{%
A}_{\mu }.
\end{eqnarray}

Finally, we find

\begin{eqnarray}
\widehat{F}_{\mu \nu } &=&F_{\mu \nu }+\theta ^{\rho \sigma }\left( x\right)
\left( F_{\mu \rho }F_{\nu \sigma }-A_{\rho }\partial _{\sigma }F_{\mu \nu
}\right)  \notag \\
&&-\frac{1}{2}\partial _{\mu }\theta ^{\rho \sigma }\left( x\right) \left(
A_{\rho }F_{\sigma \nu }+A_{\rho }\partial _{\sigma }A_{\nu }\right) \\
&&+\frac{1}{2}\partial _{\nu }\theta ^{\rho \sigma }\left( x\right) \left(
A_{\rho }F_{\sigma \mu }+A_{\rho }\partial _{\sigma }A_{\mu }\right) ,
\end{eqnarray}

which we can write as

\begin{equation}
\widehat{F}_{\mu \nu }=F_{\mu \nu }+f_{\mu \nu }+o\left( \eta ^{2}\right) ,
\end{equation}

where $f_{\mu \nu }$ is the quantum correction linear in $\eta $. The
quantum analogue of Equ. (137) is given by

\begin{equation}
\widehat{S}=-\frac{1}{4}\int d^{4}x\,\,\widehat{F}_{\mu \nu }\star \widehat{F%
}^{\mu \nu }.
\end{equation}

We can easily see from this equation that the $q$-deformed action contains
non-renormalizable vertices of dimension six. Other term which are
proportional to $\partial _{\mu }\theta ^{\rho \sigma }\left( x\right) $
appear. \newline
Finally, let us emphasize once more that we can also consider a quantum
gauge theory with a quantum gauge group as a symmetry group defined on a
quantum space. This gives a general map between deformed and ordinary gauge
fields \cite{mesref5}.

\section{The Quantum Anti-de Sitter Space}

In this section, we propose to construct the metrics of the quantum analogue
of the classical anti-de Sitter space AdS$_{5}$. \newline
The AdS$_{5}$ space is a 5-dimensional manifold with constant curvature and
signature (+,-,-,-,-). It can be embedded \ as an hyperboloid into a
6-dimensional flat space with signature (+,+,-,-,-,-), by

\begin{equation}
z_{0}^{2}+z_{5}^{2}-z_{1}^{2}-z_{2}^{2}-z_{3}^{2}-z_{4}^{2}=R^{2},
\end{equation}

where $R$ will be called the ''radius'' of the AdS$_{5}$ space. \newline

To define the quantum anti-de Sitter space we follow the method of Ref. \cite%
{chang} used for AdS$_{4}^{q}$. The quantum anti-de Sitter space AdS$^{q}
_{5}$ is nothing but the quantum sphere $S^{q} _{5}$ with a suitable reality
structure. For $\left| q\right| =1$ we consider the conjugation\footnote{%
Let us recall that a $\ast $-structure or $\ast $-conjugation on a Hopf
algebra $\mathcal{A}$ is an anti-automorphism $\left( \eta ab\right) ^{\ast
}=\bar{\eta }b^{\ast }a^{\ast } \quad \forall a,b \in \mathcal{A}, \quad
\forall \eta \in \mathbb{C}$; coalgebra automorphism $\Delta \circ \ast
=\left( \ast \otimes \ast \right) \circ \Delta $, \quad $\epsilon \circ \ast
=\epsilon $ and involution $\ast ^{2}=id$. It follows that $\ast \circ S
^{-1}=S\circ \ast $ i.e. $\left[ S ^{-1}\left( M^{n} _{\,\,\,m}\right) %
\right] ^{\ast }=S\left( M^{\ast \,n} _{\,\,\,\,\,\,\,\,\,m}\right). $}\cite%
{faddeev1} defined as $M^{\times }=M$. The unique associated quantum space
conjugation is $\left( x^{a}\right) ^{\times }=x^{a}$. By this conjugation
on the quantum orthogonal we cannot get the desired quantum AdS space. We
introduce another operation on the quantum orthogonal group as

\begin{equation}
M^{\dagger }=DMD^{-1}
\end{equation}

where the matrix $D$ is given by

\begin{equation}
D=\left( 
\begin{array}{cccccc}
1 &  &  &  &  &  \\ 
& 1 &  &  &  &  \\ 
&  & -1 &  &  &  \\ 
&  &  & -1 &  &  \\ 
&  &  &  & 1 &  \\ 
&  &  &  &  & 1%
\end{array}
\right) .
\end{equation}

We can easily prove that the $D$ matrix is a special element of the quantum
orthogonal group \cite{aschieri}. The quantum AdS group is obtained by the
combined operation $M^{\ast }\equiv M^{\times \dagger }=DMD^{-1}$. The
induced conjugation on the quantum space is $x^{\ast }\equiv x^{\times
\dagger }=Dx$. We can check that the conjugation really gives the quantum
AdS group and quantum AdS space. We should find a linear transformation $%
x\rightarrow x^{\prime }=Ux,\quad M\rightarrow M^{\prime }=UMU^{-1}$ such
that the new coordinates $x^{\prime }$ and $M^{\prime }$ are real and the
new metric $C^{\prime }=\left( U^{-1}\right) ^{t}CU^{-1}$ diagonal in the $%
q\rightarrow 1$ limit, $C^{^{\prime }}\arrowvert_{q=1}=diag\left(
1,-1,-1,-1,-1,1\right) $. The metric $C$ is defined in Equ. (45). We find
the following $U$ matrix

\begin{equation}
U=\frac{1}{\sqrt{2}}\left( 
\begin{array}{cccccc}
1 & 0 & 0 & 0 & 0 & 1 \\ 
-1 & 0 & 0 & 0 & 0 & 1 \\ 
0 & 0 & 1 & -1 & 0 & 0 \\ 
0 & 0 & i & i & 0 & 0 \\ 
0 & -1 & 0 & 0 & 1 & 0 \\ 
0 & 1 & 0 & 0 & 1 & 0%
\end{array}
\right)
\end{equation}

and the corresponding quantum metric is

\begin{equation}
C^{^{\prime }}=\left( 
\begin{array}{cccccc}
\frac{1}{2}q^{2}+\frac{1}{2q^{2}} & -\frac{1}{2}q^{2}+\frac{1}{2q^{2}} & 0 & 
0 & 0 & 0 \\ 
\frac{1}{2}q^{2}-\frac{1}{2q^{2}} & -\frac{1}{2}q^{2}-\frac{1}{2q^{2}} & 0 & 
0 & 0 & 0 \\ 
0 & 0 & -1 & 0 & 0 & 0 \\ 
0 & 0 & 0 & -1 & 0 & 0 \\ 
0 & 0 & 0 & 0 & -\frac{1}{2}q-\frac{1}{2q} & \frac{1}{2}q-\frac{1}{2q} \\ 
0 & 0 & 0 & 0 & -\frac{1}{2}q+\frac{1}{2q} & \frac{1}{2}q+\frac{1}{2q}%
\end{array}
\right) .
\end{equation}

For $q$ real we consider the second conjugation given in \cite{faddeev1} and
realized via the metric, i.e. $M^{\ast }=C^{t}MC^{t}$. The metric $C$ is
defined in Equ. (45). The condition on the braiding $R$ matrix is: $%
\overline{R}=R$. To get the quantum AdS group and the AdS space we have to
consider another operation on the quantum orthogonal space as:

\begin{equation}
M^{\ddagger }=AMA^{-1}
\end{equation}

where the matrix $A$ is given by

\begin{equation}
A=\left( 
\begin{array}{cccccc}
1 &  &  &  &  &  \\ 
& -1 &  &  &  &  \\ 
&  & -1 &  &  &  \\ 
&  &  & -1 &  &  \\ 
&  &  &  & -1 &  \\ 
&  &  &  &  & 1%
\end{array}
\right) .
\end{equation}

We obtain the AdS quantum group by using the conjugation $M^{\ast \ddagger
}=AC^{t}MC^{t}A^{-1}$. The induced conjugation on the quantum space is $%
x^{\ast \ddagger }=C^{t}Ax$. To prove that this combination really gives the
quantum AdS group and quantum AdS space we should find a linear
transformation $x\rightarrow x^{\prime }=Vx,\quad M\rightarrow M^{\prime
}=VMV^{-1}$. Such that the new coordinates $x^{\prime }$ and $M^{\prime }$
are real and the new metric $C^{^{\prime }}\arrowvert_{q=1}=diag\left(
1,-1,-1,-1,-1,1\right) $. We find the following $V$ matrix

\begin{equation}
V=\frac{1}{\sqrt{2}}\left( 
\begin{array}{cccccc}
1 & 0 & 0 & 0 & 0 & q^{2} \\ 
0 & -i & 0 & 0 & -iq & 0 \\ 
0 & 0 & 1 & -1 & 0 & 0 \\ 
0 & 0 & i & i & 0 & 0 \\ 
0 & q^{-1} & 0 & 0 & -1 & 0 \\ 
iq^{-2} & 0 & 0 & 0 & 0 & -i%
\end{array}
\right)
\end{equation}

and the quantum metric $C^{\prime }$ is given by

\begin{equation}
C^{^{\prime }}=\left( 
\begin{array}{cccccc}
\frac{1}{2}+\frac{1}{2q^{4}} & 0 & 0 & 0 & 0 & -\frac{1}{2}iq^{2}+\frac{1}{2}%
\frac{i}{q^{2}} \\ 
0 & -\frac{1}{2}-\frac{1}{2q^{2}} & 0 & 0 & \frac{1}{2}iq-\frac{1}{2}\frac{i%
}{q} & 0 \\ 
0 & 0 & -1 & 0 & 0 & 0 \\ 
0 & 0 & 0 & -1 & 0 & 0 \\ 
0 & -\frac{1}{2}iq+\frac{1}{2}\frac{i}{q} & 0 & 0 & -\frac{1}{2}-\frac{1}{2}%
q^{2} & 0 \\ 
\frac{1}{2}iq^{2}-\frac{1}{2}\frac{i}{q^{2}} & 0 & 0 & 0 & 0 & \frac{1}{2}%
q^{4}+\frac{1}{2}%
\end{array}
\right) .
\end{equation}

These metrics can be used in the definition of Lagrangians defined on the
quantum Anti-de Sitter space AdS$^{q} _{5}$. As an example of such a theory
is the quantum Chern-Simons term which is present in the low energy
effective action of type IIB superstring theory compactified on the quantum
anti-de Sitter space.

\section{$q$-deformed conformal correlation functions}

In this section, we construct the $q$-deformed two- and three- point
conformal correlation functions in field theories that are assumed to
possess an invariance under a quantum deformation of $SO\left( 4,2\right) $.
In the course of these investigations we rely on the general formalism
developed by Dobrev \cite{dobrev1} who first introduced the $q$-deformation
of $D=4$ conformal algebra \footnote{%
This quantum algebra was also studied, and in addition its contraction to
deformed Poincaré algebra given in Ref. \cite{lukierski}.} and constructed
its $q$-difference realizations. Let us recall that the positive energy
irreducible representations of $so\left( 4,2\right) $ are labelled by the
lowest value of the energy $E_{0}$, the spin $s_{0}=j_{1}+j_{2}$ and by the
helicity $h_{0}=j_{1}-j_{2}$, and these are eigenvalues of a Cartan
subalgebra $\mathcal{H}$ of $so\left( 4,2\right) $. We shall label the
representations of $U_{q}\left( so\left( 4,2\right) \right) $ in the same
way and thus we shall take for $U_{q}\left( so\left( 4,2\right) \right) $
and its complexification $U_{q}\left( so\left( 6,\mathbb{C}\right) \right) $
the same Cartan subalgebra. We recall that the $q$-deformation $U_{q}\left(
so\left( 6,\mathbb{C}\right) \right) $ is defined \cite{drinfeld,jimbo} as
the associative algebra over $\mathbb{C}$ with Chevalley generators $%
X_{j}^{\pm }$, $H_{j}$, $j=1,2,3.$

The Cartan-Chevalley basis of $U_{q}\left( sl\left( 4,\mathbb{C}\right)
\right) $ is given by the formulae:

\begin{eqnarray}
\left[ H_{j},H_{k}\right] &=&0\qquad \left[ H_{j},X_{k}^{\pm }\right] =\pm
\,a_{jk}\,X_{k}^{\pm }\qquad \qquad \qquad \qquad \qquad  \notag \\
\left[ X_{j}^{+},X_{k}^{-}\right] &=&\delta _{jk}\frac{q^{H_{j}}-q^{-H_{j}}}{%
q-q^{-1}}=\delta _{jk}\left[ H_{j}\right] _{q} .
\end{eqnarray}

and the $q$-analogue of the Serre relations

\begin{equation}
\left( X_{j}^{\pm }\right) ^{2}X_{k}^{\pm }-\left[ 2\right] _{q}X_{j}^{\pm
}X_{k}^{\pm }X_{j}^{\pm }+X_{k}^{\pm }\left( X_{j}^{\pm }\right) ^{2}=0,
\end{equation}

where $\left( jk\right) =\left( 12\right) ,\left( 21\right) ,\left(
23\right) ,\left( 32\right) $ and $\left( a_{jk}\right) $ is the Cartan
matrix of $so\left( 6,\mathbb{C}\right) $ given by $\left( a_{jk}\right)
=2\left( \alpha _{j},\alpha _{k}\right) /\left( \alpha _{j},\alpha
_{j}\right) $; $\alpha _{1},\alpha _{2},\alpha _{3}$ are the simple roots of
length 2 and the non-zero product between the simple roots are: $\left(
\alpha _{1},\alpha _{2}\right) =\left( \alpha _{2},\alpha _{3}\right) =-1$ .
The quantum number is defined as $\left[ m\right] _{q}=\frac{q^{m}-q^{-m}}{%
q-q^{-1}}$.

Explicitly the Cartan matrix is given by:

\begin{equation}
\left( a_{jk}\right) =\left( 
\begin{array}{ccc}
2 & -1 & 0 \\ 
-1 & 2 & -1 \\ 
0 & -1 & 2%
\end{array}
\right) .
\end{equation}

The elements $H_{j}$ span the Cartan subalgebra $\mathcal{H}$ while the
elements $X_{j}^{\pm }$ generate the subalgebra $\mathcal{G}^{\pm }$ in the
standard decomposition $\mathcal{G}\equiv so\left( 6,\mathbb{C}\right) =%
\mathcal{G}^{+}\oplus \mathcal{H}\oplus \mathcal{G}^{-}$. In particular, the
Cartan-Weyl generators for the non-simple roots are given by \cite{dobrev2}:

\begin{eqnarray}
X_{jk}^{\pm } &=&\pm q^{\mp 1/2}\left( q^{1/2}X_{j}^{\pm }X_{k}^{\pm
}-q^{-1/2}X_{k}^{\pm }X_{j}^{\pm }\right) \qquad \left( jk\right) =\left(
12\right) ,\left( 23\right)  \notag \\
X_{13}^{\pm } &=&\pm q^{\mp 1/2}\left( q^{1/2}X_{1}^{\pm }X_{23}^{\pm
}-q^{-1/2}X_{23}^{\pm }X_{1}^{\pm }\right)  \notag \\
&=&\pm q^{\mp 1/2}\left( q^{1/2}X_{12}^{\pm }X_{3}^{\pm }-q^{-1/2}X_{3}^{\pm
}X_{12}^{\pm }\right) .
\end{eqnarray}

All other commutation relations follow from these definitions:\quad 
\begin{eqnarray}
\left[ X_{a}^{+},X_{ab}^{-}\right] &=&-q^{H_{a}}\,X_{a+1b}^{-}\qquad \left[
X_{b}^{+},X_{ab}^{-}\right] =X_{ab-1}^{-}\,q^{-H_{b}}\quad 1\leq a<b\leq 3 
\notag \\
\left[ X_{a}^{-},X_{ab}^{+}\right] &=&X_{a+1b}^{+}\,q^{-H_{a}}\qquad \left[
X_{b}^{-},X_{ab}^{+}\right] =-q^{H_{b}}\,X_{ab-1}^{+}\quad 1\leq a<b\leq 3 
\notag \\
X_{a}^{\pm }X_{ab}^{\pm } &=&qX_{ab}^{\pm }X_{a}^{\pm }\quad \quad \quad
\quad \,\,\,X_{b}^{\pm }X_{ab}^{\pm }=q^{-1}X_{ab}^{\pm }X_{a}^{\pm }\quad
\,\,1\leq a<b\leq 3  \notag \\
X_{12}^{\pm }X_{13}^{\pm } &=&qX_{13}^{\pm }X_{12}^{\pm }\quad \quad \quad
\,\,\,\,X_{23}^{\pm }X_{13}^{\pm }=q^{-1}X_{13}^{\pm }X_{23}^{\pm }  \notag
\\
\left[ X_{2}^{\pm },X_{13}^{\pm }\right] &=&0\qquad \quad \quad \quad \quad
\,\,\left[ X_{2}^{\pm },X_{13}^{\mp }\right] =0\qquad  \notag \\
\left[ X_{12}^{+},X_{13}^{-}\right] &=&-q^{2\left( H_{1}+H_{2}\right)
}X_{3}^{-}\quad \left[ X_{12}^{-},X_{13}^{+}\right] =X_{3}^{+}q^{-2\left(
H_{1}+H_{2}\right) }  \notag \\
\qquad \left[ X_{23}^{+},X_{13}^{-}\right] &=&X_{1}^{-}\,q^{-2\left(
H_{2}+H_{3}\right) }\quad \left[ X_{23}^{-},X_{13}^{+}\right] =-q^{2\left(
H_{2}+H_{3}\right) }X_{1}^{+}\,  \notag \\
\left[ X_{12}^{\pm },X_{23}^{\pm }\right] &=&\lambda X_{2}^{\pm }X_{13}^{\pm
}\quad \quad \quad \,\left[ X_{12}^{\pm },X_{23}^{\mp }\right] =-\lambda
q^{\pm H_{2}}X_{1}^{\pm }X_{3}^{\mp }
\end{eqnarray}

where $\lambda =q-q^{-1}$.

The dilatation generator is given by

\begin{equation}
D =\frac{1}{2}\left( H_{1}+H_{3}\right) +H_{2} .
\end{equation}

The quantum universal algebra $U_{q}\left( su\left( 2,2\right) \right) $ is
a Hopf algebra with coproduct defined by:

\begin{eqnarray}
\Delta \left( H_{i}\right) &=&H_{i}\otimes 1+1\otimes H_{i}  \notag \\
\Delta \left( X_{\pm i}\right) &=&X_{\pm i}\otimes
q^{H_{i}/2}+q^{-H_{i}/2}\otimes X_{\pm i}.
\end{eqnarray}

and antipode and counit defined as

\begin{eqnarray}
S\left( H_{i}\right) &=&-H_{i},  \notag \\
S\left( X_{+i}\right) &=&-qX_{+i},\qquad S\left( X_{-i}\right)
=-q^{-1}X_{-i},  \notag \\
\epsilon \left( H_{i}\right) &=&\epsilon \left( X_{+i}\right) =0.
\end{eqnarray}

Now, let us compute the $q$-deformed 2-point conformal correlation function
of scalar quasiprimary (qp) fields, with canonical dimension $d_{1}$ and $%
d_{2}$, defined on the $q$-deformed Minkowski spacetime \footnote{%
Up to Equ. (184) this section follows the paper \cite{dobrev3}.}\cite%
{dobrev3}.

\begin{eqnarray}
x_{\pm }v &=&q^{\pm 1}vx_{\pm },\quad \quad \quad \quad \qquad \,\,\, \qquad
x_{\pm }\overline{v}=q^{\pm 1}\overline{v}x_{\pm },  \notag \\
\lambda v\overline{v} &=&x_{+}x_{-}-x_{-}x_{+},\quad \,\, \quad \qquad
\qquad \overline{v}v=v\overline{v},  \notag \\
x_{\pm } &\equiv &x^{0}\pm x^{3}\quad \,\,\,\,\quad v\equiv
x^{1}-ix^{2}\quad \quad \overline{v}\equiv x^{1}+ix^{2}.
\end{eqnarray}

The $q$-Minkowski length is

\begin{equation}
\mathcal{L}_{q}=x_{-}x_{+}-q^{-1}v\overline{v}.
\end{equation}

These qp-fields are reduced functions and can be written as formal power
series in the $q$-Minkowski coordinates:

\begin{eqnarray}
\phi &=&\phi \left( Y\right) =\phi \left( v,x_{-},x_{+},\overline{v}\right) 
\notag \\
&=&\sum_{j,n,l,m\in Z_{+}}\mu _{jnlm}\,\phi _{j\,\,n\,\,l\,\,m},  \notag \\
\phi _{jnlm} &=&v^{j}x_{-}^{n}x_{+}^{l}\overline{v}^{m}.
\end{eqnarray}

Next we introduce the following operators acting on the reduced functions as 
\begin{eqnarray}
\widehat{M}_{\kappa }\phi \left( Y\right) &=&\sum_{j,n,l,m\in Z_{+}}\mu
_{jnlm}\,\widehat{M}_{\kappa }\,\phi _{j\,\,n\,\,l\,\,m}  \notag \\
T_{\kappa }\phi \left( Y\right) &=&\sum_{j,n,l,m\in Z_{+}}\mu
_{jnlm}T_{\kappa }\phi _{j\,\,n\,\,l\,\,m}
\end{eqnarray}

where $\kappa =\pm ,v,\overline{v}$ and the explicit action on $\phi
_{j\,\,n\,\,l\,\,m}$ is defined by

\begin{eqnarray}
\widehat{M}_{v}\phi _{j\,\,n\,\,l\,\,m} &=&\phi _{j+1\,\,n\,\,l\,\,m}  \notag
\\
\widehat{M}_{-}\phi _{j\,\,n\,\,l\,\,m} &=&\phi _{j\,\,n+1\,\,l\,\,m}  \notag
\\
\widehat{M}_{+}\phi _{j\,\,n\,\,l\,\,m} &=&\phi _{j\,\,n\,\,l+1\,\,m}  \notag
\\
\widehat{M}_{\overline{v}}\phi _{j\,\,n\,\,l\,\,m} &=&\phi
_{j\,\,n\,\,l\,\,m+1}  \notag \\
T_{v}\phi _{j\,\,n\,\,l\,\,m} &=&q^{j}\phi _{j\,\,n\,\,l\,\,m}  \notag \\
T_{-}\phi _{j\,\,n\,\,l\,\,m} &=&q^{n}\phi _{j\,\,n\,\,l\,\,m}  \notag \\
T_{+}\phi _{j\,\,n\,\,l\,\,m} &=&q^{l}\phi _{j\,\,n\,\,l\,\,m}  \notag \\
T_{\overline{v}}\phi _{j\,\,n\,\,l\,\,m} &=&q^{m}\phi _{j\,\,n\,\,l\,\,m}.
\end{eqnarray}

The $q$-difference operators are defined by

\begin{equation}
\widehat{D}_{\kappa }\phi =\frac{1}{\lambda }\widehat{M}_{\kappa
}^{-1}\left( T_{\kappa }-T_{\kappa }^{-1}\right) \phi .
\end{equation}

\newpage

The representation action of $U_{q}\left( sl\left( 4\right) \right) $ on the
reduced functions $\phi \left( Y\right) $ of the representation space $%
C^{\Lambda }$, with the signature $\chi =\chi \left( \Lambda \right) =\left(
m_{1},m_{2},m_{3}\right) =\left( 1,1-d,1\right) $ and which corresponds to a
spinless \textquotedblleft scalar\textquotedblright\ field $\left[
d,j_{1},j_{2}\right] =\left[ d,0,0\right] $ is given by\footnote{%
The general case is given in Ref. \cite{dobrev2}} :

\begin{eqnarray}
\pi \left( k_{1}\right) \phi _{j\,\,n\,\,l\,\,m} &=&q^{\left( j-n+l-m\right)
/2}\phi _{j\,\,n\,\,l\,\,m},  \notag \\
\pi \left( k_{2}\right) \phi _{j\,\,n\,\,l\,\,m} &=&q^{n+\left( j+m+d\right)
/2}\phi _{j\,\,n\,\,l\,\,m},  \notag \\
\pi \left( k_{3}\right) \phi _{j\,\,n\,\,l\,\,m} &=&q^{\left(
-j-n+l+m\right) /2}\phi _{j\,\,n\,\,l\,\,m},  \notag \\
\pi \left( X_{+1}\right) \phi _{j\,\,n\,\,l\,\,m} &=&q^{-1+\left(
j-n-l+m\right) /2}\left[ n\right] _{q}\phi _{j+1\,\,n-1\,\,l\,\,m}  \notag \\
&&+q^{-1+\left( j-n+l-m\right) /2}\left[ m\right] _{q}\phi
_{j\,\,n\,\,l+1\,\,m-1},  \notag \\
\pi \left( X_{+2}\right) \phi _{j\,\,n\,\,l\,\,m} &=&q^{\left( -j+m\right)
/2}\left[ j+n+m+d\right] _{q}\phi _{j\,\,n+1\,\,l\,\,m}  \notag \\
&&+q^{d+\left( j+n+3m\right) /2}\left[ l\right] _{q}\phi
_{j+1\,\,n\,\,l-1\,\,m+1},  \notag \\
\pi \left( X_{+3}\right) \phi _{j\,\,n\,\,l\,\,m} &=&-q^{-1+\left(
j+n-l-m\right) /2}\left[ j\right] _{q}\phi _{j-1\,\,n\,\,l+1\,\,m}  \notag \\
&&-q^{-1+\left( 3j+n-3l-m\right) /2}\left[ n\right] _{q}\phi
_{j\,\,n-1\,\,l\,\,m+1},  \notag
\end{eqnarray}

\begin{eqnarray}
\pi \left( X_{-1}\right) \phi _{j\,\,n\,\,l\,\,m} &=&q^{2+\left(
-j+n-l+m\right) /2}\left[ j\right] _{q}\phi _{j-1\,\,n+1\,\,l\,\,m}  \notag
\\
&&+q^{2+\left( j-n-l+m\right) /2}\left[ l\right] _{q}\phi
_{j\,\,n\,\,l\,-1\,m+1},  \notag \\
\pi \left( X_{-2}\right) \phi _{j\,\,n\,\,l\,\,m} &=&-q^{\left( j-m\right)
/2}\left[ n\right] _{q}\phi _{j\,\,n-1\,\,l\,\,m},  \notag \\
\pi \left( X_{-3}\right) \phi _{j\,\,n\,\,l\,\,m} &=&-q^{\left(
-j-3n+l+3m\right) /2}\left[ l\right] _{q}\phi _{j+1\,\,n\,\,l-1\,\,m}  \notag
\\
&&-q^{\left( -j-n+l+m\right) /2}\left[ m\right] _{q}\phi
_{j\,\,n+1\,\,l\,\,m-1},
\end{eqnarray}

with $k_{i}=q^{H_{i}/2}$.

Now let us define $\mathcal{D}=q^{D}$, where $D$ is the dilatation generator
defined in Equ. (173). The representation of this generator on the reduced
functions $\phi $ is given by

\begin{eqnarray}
\pi \left( \mathcal{D}\right) \phi \left( Y\right) &=&\mu _{jnlm}\pi \left( 
\mathcal{D}\right) \phi _{j\,\,n\,\,l\,\,m}  \notag \\
\, &=&q^{d}\mu _{jnlm}q^{j+n+l+m}\phi _{j\,\,n\,\,l\,\,m}=q^{d}\phi \left(
qY\right) .
\end{eqnarray}

The coproduct for this operator is given by

\begin{equation}
\Delta \mathcal{D}=\mathcal{D}\otimes \mathcal{D}\text{.}
\end{equation}

Now let us calculate two point $q$-correlation functions by imposing that
they are invariant under the action of $U_{q}\left( sl\left( 4,\mathbb{C}
\right) \right) $. We denote the $q$-deformed correlation functions of $N$
quasiprimary fields as

\begin{equation}
\left\langle \phi _{1}\left( Y_{1}\right) ...\phi _{N}\left( Y_{N}\right)
\right\rangle _{q}=\,_{q}\left\langle 0\left\vert \phi _{d_{1}}\left(
Y_{1}\right) ...\phi _{d_{N}}\left( Y_{N}\right) \right\vert 0\right\rangle
_{q},
\end{equation}

where $\arrowvert0\rangle _{q}$ is a $U_{q}\left( sl\left( 4,\mathbb{C}%
\right) \right) $ invariant vacuum such that $\pi \left( \mathcal{D}\right) $%
$\arrowvert0\rangle _{q}=\arrowvert0\rangle _{q}$, $\pi \left( X_{+i}\right) %
\arrowvert0\rangle _{q}=0$ and also for $_{q}\langle 0\arrowvert$ . The
identities for the two-point correlation functions of two quasiprimary
fields of the conformal weights $d_{1}$, $d_{2}$ are

\begin{eqnarray}
\Delta \left( \pi \left( \mathcal{D}\right) \right) \left\langle \phi
_{1}\left( Y_{1}\right) \phi _{2}\left( Y_{2}\right) \right\rangle _{q}
&=&\left( \pi \left( \mathcal{D}\right) \otimes \pi \left( \mathcal{D}%
\right) \right) \left\langle \phi _{1}\left( Y_{1}\right) \phi _{2}\left(
Y_{2}\right) \right\rangle  \notag \\
&=&\left\langle \phi _{1}\left( Y_{1}\right) \phi _{2}\left( Y_{2}\right)
\right\rangle _{q}
\end{eqnarray}

and

\begin{eqnarray}
&&\Delta \left( \pi \left( X_{\pm i}\right) \right) \left\langle \phi
_{1}\left( Y_{1}\right) \phi _{2}\left( Y_{2}\right) \right\rangle _{q}= 
\notag \\
&&\left( \pi \left( X_{\pm i}\right) \otimes q^{\pi \left( H_{i}/2\right)
}+q^{-\pi \left( H_{i}/2\right) }\otimes \pi \left( X_{\pm i}\right) \right)
.  \notag \\
&&\left\langle \phi _{1}\left( Y_{1}\right) \phi _{2}\left( Y_{2}\right)
\right\rangle _{q}=0.
\end{eqnarray}

The $q$-correlation functions are covariant under dilatation, whereas the
remaining identities lead to six $q$-difference equations.

Let us first note that

\begin{eqnarray}
\phi _{j+1\,\,n-1\,\,l\,\,m} &=&q^{j}v\left( x_{-}\right) ^{-1}\,\,\phi
_{j\,\,n\,\,l\,\,m},\quad  \notag \\
\phi _{j\,\,n\,\,l+1\,\,m-1} &=&q^{m}\phi _{j\,\,n\,\,l\,\,m}\,\,x_{+}\left( 
\overline{v}\right) ^{-1},
\end{eqnarray}

and so on,

\begin{eqnarray}
q^{\pm j/2}\phi \left( v,x_{-},x_{+},\overline{v}\right) &=&\phi \left(
q^{\pm 1/2}v,x_{-},x_{+},\overline{v}\right) ,  \notag \\
q^{\pm n/2}\phi \left( v,x_{-},x_{+},\overline{v}\right) &=&\phi \left(
v,q^{\pm 1/2}x_{-},x_{+},\overline{v}\right) ,...
\end{eqnarray}

and

\begin{eqnarray}
\left[ n\right] _{q}\phi &=&\lambda ^{-1}\left( \phi \left( v,qx_{-},x_{+},%
\overline{v}\right) -\phi \left( v,q^{-1}x_{-},x_{+},\overline{v}\right)
\right)  \notag \\
&=& \widehat{D}_{-}\,\phi \left( v,x_{-},x_{+},\overline{v}\right) ,  \notag
\\
\left[ m\right] _{q}\phi &=&\lambda ^{-1}\left( \phi \left( v,x_{-},x_{+},q%
\overline{v}\right) -\phi \left( v,x_{-},x_{+},q^{-1}\overline{v}\right)
\right)  \notag \\
&=& \widehat{D}_{\overline{v}}\,\phi \left( v,x_{-},x_{+},\overline{v}%
\right) ,...
\end{eqnarray}

and so forth.

The first identity for $X_{+1}$ is given by:

\begin{eqnarray}
&&q^{j}v_{1}\left( x_{-1}\right) ^{-1}\,\,\langle \widehat{D}_{-}\,\phi
_{1}\left( q^{1/2}v_{1},q^{-1/2}x_{-1},q^{-1/2}x_{+1},q^{1/2}\overline{v}%
_{1}\right)  \notag \\
&&\times \,\, \phi _{2}\left(
q^{1/2}v_{2},q^{-1/2}x_{-2},q^{1/2}x_{+2},q^{-1/2}\overline{v}_{2}\right)
\rangle _{q}  \notag \\
&&+ \,\, q^{m}\langle \widehat{D}_{\overline{v}}\phi _{1}\left(
q^{1/2}v_{1},q^{-1/2}x_{-1},q^{1/2}x_{+1},q^{-1/2}\overline{v}_{1}\right)
x_{+1}\left( \overline{v}_{2}\right) ^{-1}  \notag \\
&&\times \,\, \phi _{2}\left(
q^{1/2}v_{2},q^{-1/2}x_{-2},q^{1/2}x_{+2},q^{-1/2}\overline{v}_{2}\right)
\rangle _{q}  \notag \\
&&+ \,\, q^{j}v_{2}\left( x_{-2}\right) ^{-1}\,\,\langle \phi _{1}\left(
q^{1/2}v_{1},q^{-1/2}x_{-1},q^{1/2}x_{+1},q^{-1/2}\overline{v}_{-1}\right) 
\notag \\
&&\times \,\, \widehat{D}_{-}\phi _{2}\left(
q^{1/2}v_{2},q^{-1/2}x_{-2},q^{-1/2}x_{+2},q^{1/2}\overline{v}_{2}\right)
\rangle _{q}  \notag \\
&&+ \, \,\, q^{m}\langle \phi _{1}\left(
q^{1/2}v_{1},q^{-1/2}x_{-1},q^{1/2}x_{+1},q^{-1/2}\overline{v}_{1}\right)
x_{+2}\left( \overline{v}_{2}\right) ^{-1}  \notag \\
&&\times \,\, \widehat{D}_{\overline{v}}\phi _{2}\left(
q^{1/2}v_{2},q^{-1/2}x_{-2},q^{1/2}x_{+2},q^{-1/2}\overline{v}_{1}\right)
\rangle _{q}=0
\end{eqnarray}

and five other $q$-difference equations.

The solution of these $q$-difference equations exists when the conformal
dimensions $d_{1}$ and $d_{2}$ are equal: $d_{1}=d_{2}=d$ and is determined
uniquely up to a constant. Let us use the twistors $Y=Y ^{\mu }\sigma _{\mu
} $. More explicitly, the matrices $Y$ are given by:

\begin{equation}
Y=\left( 
\begin{array}{cc}
x_{0}+x_{3} & x_{0}-ix_{2} \\ 
x_{0}+ix_{2} & x_{0}-x_{3}%
\end{array}%
\right) =\left( 
\begin{array}{cc}
x_{+} & v \\ 
\overline{v} & x_{-}%
\end{array}%
\right) ,
\end{equation}

where $x_{+},x_{-},v,\overline{v}$ are $q$-deformed Minkowski coordinates
defined in Equ. (176).

It is easy to see that the quantum determinant\footnote{%
We use Manin's notation \cite{manin}.}

\begin{equation}
det_{q}Y=x_{-}x_{+}-q^{-1}v\overline{v}\quad \text{and}\quad det_{q}\left(
Y_{1}-Y_{2}\right) =det_{q}\,Y_{1}\left( I-Y_{1}^{-1}Y_{2}\right) ,
\end{equation}

where $I$ is a $2\times 2$ identity matrix. The $q$-deformed two-point
conformal correlation function reads

\begin{equation}
\left\langle \phi _{1}\left( Y_{1}\right) \phi _{2}\left( Y_{2}\right)
\right\rangle _{q}=C\left( q\right) \,\,\left( det_{q}\,Y_{1}\right)
^{-d}\,\,\,\,\,_{1}\varphi _{0}\left( d;q^{1-d/2}Y_{1}^{-1}Y_{2}\right) ,
\end{equation}

where $C\left( q \right) $ is a constant and where the quantum
hypergeometric function with matricial argument\footnote{%
classical hypergeometric functions with matricial argument were introduced
by Bochner \cite{bochner} through Bessel functions with matricial argument.
They have been used in generating probability distributions as a
generalization of the use of classical hypergeometric functions \cite%
{rodriguez}.} is given by:

\begin{equation}
\,_{1}\varphi _{0}\left( d;Y\right) =det_{q}\prod_{l=0}^{\infty }\left(
I-q^{l}Y\right) ^{-1}\left( I-q^{d+l}Y\right) .
\end{equation}

One easily sees that the $q$-correlation function reduces to the undeformed
conformal correlation function because $\,_{1}\varphi _{0}\left( d;Y\right) $
becomes \newline
$\,_{1}F_{0}\left( d;Y\right) =det\left( I-Y\right) ^{-d}$ in the limit $%
q\rightarrow 1$.

The identities for the $q$-deformed three-point conformal correlation
functions read

\begin{eqnarray}
&&(\pi \left( X_{i}\right) \otimes q^{\pi \left( -H_{i}/2\right) }\otimes
q^{\pi \left( -H_{i}/2\right) }+q^{\pi \left( -H_{i}/2\right) }\otimes \pi
\left( X_{i}\right) \otimes q^{\pi \left( -H_{i}/2\right) }  \notag \\
&&+q^{\pi \left( -H_{i}/2\right) }\otimes q^{\pi \left( -H_{i}/2\right)
}\otimes \pi \left( X_{i}\right) )\left\langle \phi _{1}\left( Y_{1}\right)
\phi _{2}\left( Y_{2}\right) \phi _{3}\left( Y_{3}\right) \right\rangle
_{q}=0,
\end{eqnarray}

\begin{eqnarray}
&&\left( \pi \left( \mathcal{D}\right) \otimes \pi \left( \mathcal{D}\right)
\otimes \pi \left( \mathcal{D}\right) \right) \left\langle \phi _{1}\left(
Y_{1}\right) \phi _{2}\left( Y_{2}\right) \phi _{3}\left( Y_{3}\right)
\right\rangle _{q}=  \notag \\
&&\left\langle \phi _{1}\left( Y_{1}\right) \phi _{2}\left( Y_{2}\right)
\phi _{3}\left( Y_{3}\right) \right\rangle _{q}.
\end{eqnarray}

The solutions are given by

\begin{eqnarray}
&&\left\langle \phi _{1}\left( Y_{1}\right) \phi _{2}\left( Y_{2}\right)
\phi _{3}\left( Y_{3}\right) \right\rangle _{q}=C_{ijk}\,\,  \notag \\
&&\left( det_{q}Y_{1}\right) ^{-\gamma _{12}^{3}}\,\,_{1}\varphi _{0}\left(
\gamma _{12}^{3};q^{1-d_{1}/2}Y_{1}^{-1}Y_{2}\right) .  \notag \\
&&\left( det_{q}Y_{2}\right) ^{-\gamma _{23}^{1}}\,\,_{1}\varphi _{0}\left(
\gamma _{23}^{1};q^{1-d_{2}/2}Y_{2}^{-1}Y_{3}\right) .  \notag \\
&&\left( det_{q}Y_{1}\right) ^{-\gamma _{31}^{2}}\,\,_{1}\varphi _{0}\left(
\gamma _{31}^{2};q^{1+\frac{d_{2}-d_{1}}{2}}Y_{1}^{-1}Y_{3}\right) ,
\end{eqnarray}

where $\gamma _{ij}^{k}=\frac{d_{k}-d_{i}-d_{j}}{2}$ and $C_{ijk}$ are the
structure constants.

Let us mention once more that at $q=1$ we get the ordinary conformal
correlation functions. \newline
These results are the first steps towards the construction of the quantum
bulk to boundary and bulk to bulk propagators in the quantum AdS$_{5}^{q}/%
\mathrm{CFT}_{4}^{q}$ correspondence.

\newpage

\section{Conclusion and Outlook}

Quantum Groups emerged as generalized symmetries in the study of quantum
integrable systems. Ever since, they become the cornerstone in theoretical
physics. Recent discoveries of Vafa and his collaborators \cite{vafa}
clearly prove that $q$-deformation will play an important role in string
theory, Yang-Mills theory and Black holes. The aim of this review was to
present a recent and pedagogical overview of the definitions and methods
used in quantum groups and quantum gauge theory and highlight some recent
results found by the author. On going and envisaged work involves
investigations of the $q$-deformed string theory, not just by deforming the
oscillators, but by using the formal properties of the Hopf algebra
structures. Next, we will try to deeply understand the connection of these
new $q$-strings to Vafa's work.

\paragraph{Acknowledgements.}

I would like to extend very special thanks to the members of the High Energy
Theory Group at Brown university for warm hospitality. I am very grateful to
A. Sudbery for sending his very interesting papers. This work was partially
supported by Universit\'e des Sciences et de la Technologie d'Oran, USTOMB
under grant PFD USTOMB 04.

\end{document}